# Josephson Field Effect Transistors with InAs on Insulator and High Permittivity Gate Dielectrics

Alessandro Paghi[1*], Laura Borgongino[1], Sebastiano Battisti[1], Simone Tortorella[1,2], Giacomo Trupiano[1], Giorgio De Simoni[1], Elia Strambini[1], Lucia Sorba[1], and Francesco Giazotto[1]

[1]Istituto Nanoscienze-CNR and Scuola Normale Superiore, Piazza San Silvestro 12, 56127 Pisa, Italy.

[2]Dipartimento di Ingegneria Civile e Industriale, Università di Pisa, Largo Lucio Lazzarino, 56122 Pisa, Italy

[*]Corresponding authors: alessandro.paghi@nano.cnr.it

## Abstract

InAs on Insulator (InAsOI) has been recently demonstrated as a promising platform to develop hybrid semiconducting-superconducting Josephson Junctions (JJs) and Josephson Field Effect Transistors (JoFETs). The InAsOI consists of an InAs epilayer grown onto a cryogenic-electrically-insulating InAlAs metamorphic buffer, which allows the electrical decoupling of surface-exposed adjacent devices together with a high critical current density integration. The miniaturization of Si microchips has progressed significantly due to the integration of high permittivity (high-*k*) gate insulators, allowing an increased gate coupling with the transistor channel with consequent reduced gate operating voltages and leakages. As well as for Si-based FETs, integrating high-*k* gate insulators with JoFETs promises similar advantages in superconducting electronics. Here, we investigate the gate-tunable electrical properties of InAsOI-based JoFETs featuring different high-*k* gate insulators, namely, $HfO_2$ and $Al_2O_3$. We found that both the ungated and gate-tunable electrical properties of the JoFETs are strongly dependent on the insulator chosen. With both dielectrics, the JoFETs can entirely suppress the switching current and increase the normal state resistance by 10-20 times using negative gate voltages. The $HfO_2$-JoFETs exhibit improved gate-tunable electrical performance compared to those achieved with $Al_2O_3$-JoFETs, which is related to the higher permittivity of the insulator. Gate-dependent electrical properties of InAsOI-based JoFETs were evaluated in the temperature range from 50 mK to 1 K. As expected, the switching current monotonically decreases with the increase in temperature, while the normal state resistance remains unchanged until 1 K. Moreover, under the influence of an out-of-plane magnetic field, JoFETs exhibited an unconventional Fraunhofer diffraction pattern, from which an edge-peaked supercurrent density distribution was calculated. The origin of such anomalies is identified in the physics of the JJ edges, either with an increased current density or with a more accurate consideration of non-uniform flux focusing on the superconducting leads.

## Introduction

Hybrid superconducting-semiconducting electronics is the new star of cryogenic fast and ultra-low power consumption solid-state quantum electronics [1,2]. Over the last decades, hybrid superconducting-semiconducting devices have been involved as fundamental blocks to explore novel physical phenomena [3–6] and to build quantum electronic architectures [7,8] helpful in quantum computation.
In the III-V compound group, InAs [9–12], InGaAs [13–15], and GaAs [16,17] were employed as the semiconducting part of hybrid superconducting-semiconducting devices, where the superconducting properties can be lent via the superconducting proximity effect [18,19]. Among the morphological platforms, 2D quantum wells [9,11,15,20] and 1D nanowires [10,21] are conventionally employed to implement circuits featuring hybrid superconductor-semiconductor devices, where different kinds of superconductors, namely, Al [10,11], Pb [22], Nb [9,15], and NbTi [23,24], were used to induce different superconducting gaps.

InAs on Insulator (InAsOI) is a new heterostructure developed for hybrid semiconducting-superconducting electronics, which consists of an InAs epilayer grown onto a cryogenic-electrically-insulating InAlAs metamorphic buffer [12,25]. This scheme allows the electrical decoupling of surface-exposed adjacent devices. Compared to other InAs-based platforms, InAsOI enables manufacturing Josephson Junctions (JJs) with higher critical current densities, suitable for supercurrent multiplexing and demultiplexing [26]. The supercurrent control of InAsOI-based JJs was obtained by engineering the device architecture, including a gate electrode, realizing a Josephson Field Effect Transistor (JoFET) [9–11]. The control of the *n*-type InAs charge density with the gate voltage results in a tuning of the superconducting properties of the InAsOI-based JoFET: the higher the negative gate voltage absolute value, the lower (higher) the JoFET switching current (normal-state resistance) [26].

The JoFET is an essential building block for quantum fundamental research activities [27–29] and technological applications [30–33]. For instance, the gatemon qubit is a resonant superconducting circuit that relies on the nonlinear inductance of a JoFET [20,21,34–36]. The qubit transition frequency can be tuned by using the gate voltage to change the inductance of the JoFET itself, promising to simplify the scaling-down of multi-qubit circuits and to provide new methods to control multi-qubit architectures. InAs also features strong spin-orbit interaction, allowing the conversion of magnetic field or ferromagnetic impurities into a persistent phase bias ($\varphi_0$) across a JJ to achieve a Josephson phase battery [37]. The gate control of spin-orbit interaction [38] makes the JoFET a promising electrically-tuned $\varphi_0$-JJ.

Starting from the 45-nm technological node, the miniaturization of Si-based microchips has progressed significantly due to the integration of high permittivity (high-*k*) gate insulators, compared to the conventional thermally-growth $SiO_2$ [39–41]. This increased the gate coupling with the transistor channel, leading to the possibility of reducing gate operating voltages and leakages. As well as for Si-based FET, integrating high-*k* gate insulators with JoFETs promises similar advantages in superconducting electronics.

In this work, we report on the gate-tunable electrical properties of InAsOI-based JoFETs with different high-*k* gate insulators, namely, $HfO_2$ and $Al_2O_3$. We found that $HfO_2$-JoFETs exhibited superior abilities in tuning the switching current and the normal state resistance values with the gate voltage compared to those of $Al_2O_3$-JoFETs, which is related to the higher relative permittivity of the insulator. However, for both the dielectrics, a total suppression of the switching current together with 10 – 20 times increase of the normal-state resistance was obtained by applying negative gate voltages. We measured the fabricated JoFETs' temperature-dependent behavior in the 50 – 1000 mK range, and we observed a decrease in the switching current with the increase in temperature, regardless of the gate voltage and the gate insulator chosen. In contrast, the normal-state resistance vs. gate voltage trend remains unchanged until 1 K. Additionally, we collected the out-of-plane magnetic field diffraction patterns of the JoFETs supercurrent, from which we extracted the relationship between the supercurrent density and the applied gate voltage. Moreover, we observed deviations with respect to the conventional Fraunhofer-like diffraction that we justified with edge effects.

## Experimental Section

### *InAsOI Heterostructure Growth via Molecular Beam Epitaxy*

InAsOI was grown on semi-insulating GaAs (100) substrates using solid-source Molecular Beam Epitaxy (MBE). Starting from the GaAs substrate, the sequence of the layer structure includes a 200 nm-thick GaAs layer, a 200 nm-thick GaAs/$Al_{0.16}Ga_{0.84}As$ superlattice, a 200 nm-thick GaAs layer, a 1.250 µm-thick step-graded $In_XAl_{1-X}As$ metamorphic buffer (with X increasing from 0.15 to 0.81), a 400 nm-thick $In_{0.84}Al_{0.16}As$ overshoot layer, and a 100 nm-thick InAs layer. The metamorphic buffer and the overshoot layers were grown at optimized substrate temperatures of 320 °C ± 5 °C. The InAs epilayer was grown at 480 ± 5 °C.

### *InAsOI-based Josephson Field Effect Transistor Fabrication*

Al-InAs-Al JoFETs were fabricated with width ($W_{JJ}$) of 6 µm, interelectrode separation ($L_{JJ}$) ranging from 500 to 1250 nm, and gate lengths ($L_G$) of 200 and 1000 nm. First, InAsOI substrates were etched from native InAs oxide ($InAsO_X$) and passivated with Sulfur-termination dipping the InAsOI samples in a $(NH_4)_2S_X$ solution (290 mM $(NH_4)_2S$ and 300 mM S in $H_2O$). Then, a 100-nm-thick Al layer was deposited at a rate of 2 A/s at a residual chamber pressure of 1E-8 ÷ 5E-9 Torr. JJs were fabricated via two aligned lithographic steps: first, Al and InAs mesa were defined by UV-lithography and manufactured by successive Al and InAs wet etching, setting the JoFET width and leaving the cryogenic-dielectric-InAlAs layer exposed. Al was removed by dipping samples in the Transene Al Etchant Type D, while InAs was etched by dipping samples in a $H_3PO_4$:$H_2O_2$ solution (348 mM $H_3PO_4$, 305 mM $H_2O_2$ in $H_2O$). Then, Ti/Au cross-type markers were fabricated to align all the next steps with the JoFETs mesa. JoFETs interelectrode separations were defined by aligned-electron-beam-lithography and Al wet-etching, leaving the underneath InAs unaffected. Successively, 30 nm-thick $HfO_2$ or $Al_2O_3$ layers were deposited through 250 Atomic Layer Deposition (ALD) cycles at an optimized temperature of 130 °C and used as gate insulators. For $HfO_2$, Tetrakis(ethylmethylamino)hafnium (TEMAH) and $H_2O$ were used as oxide precursors, while Trimethylaluminum (TMAl) and $H_2O$ were involved in the $Al_2O_3$ growth. In both cases, Ar was used as carrier gas. Eventually, JoFETs metallic gates were defined by aligned-electron-beam-lithography followed by a deposition of a 6/60-nm-thick Ti/Al bilayer at a rate of 0.5/2 A/s with a tilt angle of 53° at a residual chamber pressure of 1E-8 ÷ 1E-9 Torr.

## Results and Discussion

Figure 1a shows the cross-section structure of a JoFET fabricated on InAsOI platform.

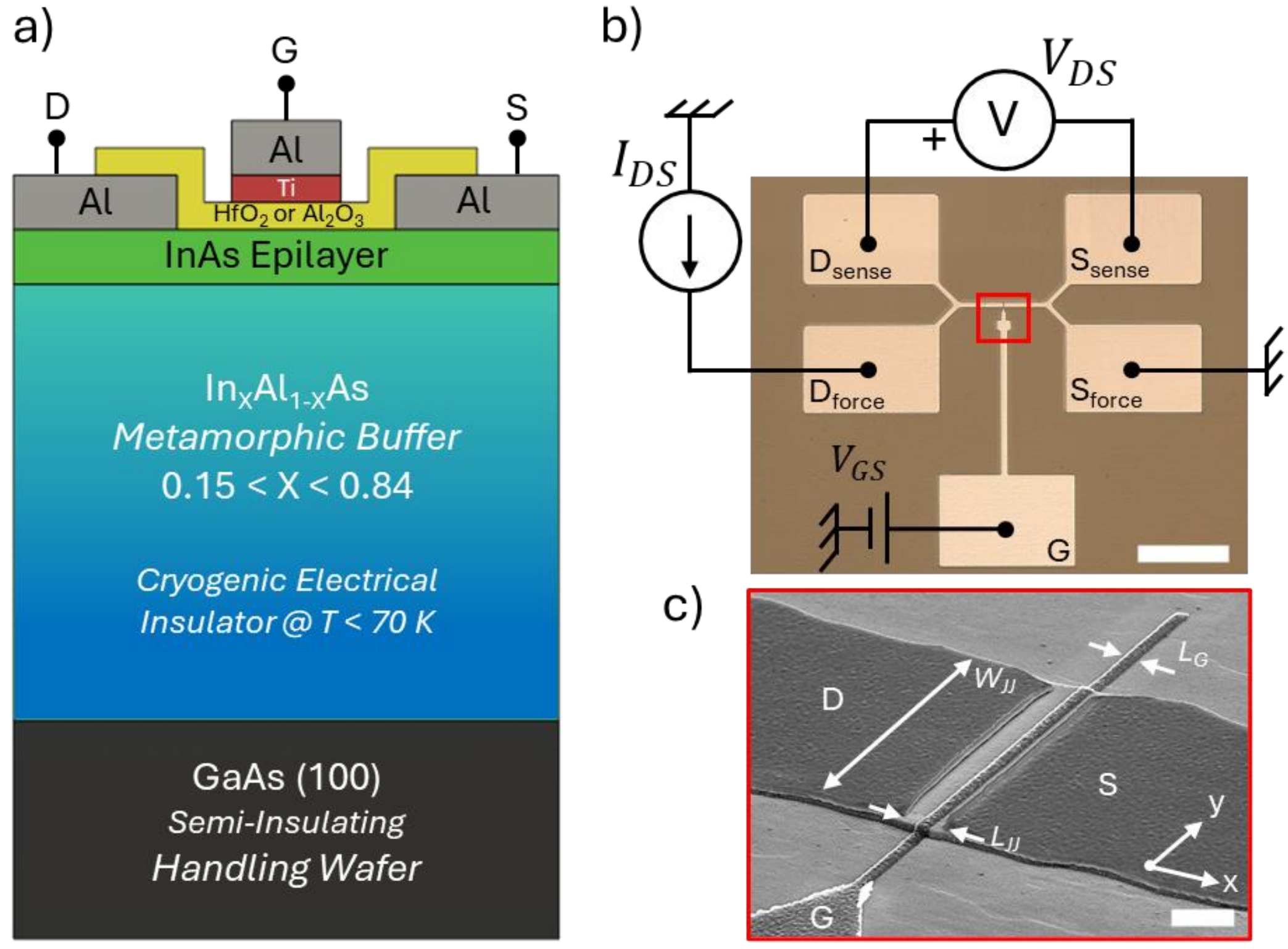

**Figure 1: Cross-section structure and morphological characterization of InAsOI-based Josephson Field Effect Transistors.** a) Cross-section structure of a JoFET with InAsOI. b) Optical microscope image of the 4-terminals JoFET architecture superimposed with the electrical measurement configuration; the scalebar is 100 μm. c) Tilted-view SEM micrograph (60°, 15k ×) of the red inset in (b) showing a JoFET with $W_{JJ}$ = 6 μm, $L_{JJ}$ = 800 nm, and $L_G$ = 200 nm; the scalebar is 1 μm.

The InAsOI stack was grown by MBE and consists of a 500 μm-thick semi-insulating GaAs (100) substrate, a ~1.5 μm-thick step-graded $In_XAl_{1-X}As$ metamorphic buffer with X increasing from 0.15 to 0.84, and finally, a 100 nm-thick intrinsically *n*-type doped semiconducting InAs epilayer. The InAs features a sheet electron density ($n_{2D}$) of $1.4\times10^{12}$ $cm^{-2}$, a sheet resistivity ($\rho_{2D}$) of 680 Ω, and a mobility ($\mu_n$) of $6.7\times10^3$ $cm^2$/Vs at 3 K. $n_{2D}$ is lower than the value measured at 300 K, namely, $5.5\times10^{12}$ $cm^{-2}$, which is related to the charge freezeout effect both in the InAs and InAlAs layers due to the temperature reduction [42–44]. Despite the semiconductive behavior of InAs, the InAlAs metamorphic buffer layer acts as an electrical insulator for temperatures lower than 70 K, allowing electrical decoupling of adjacent devices [12]. An Al layer (100 nm-thick) was used as the superconductor, a Ti/Al bilayer (6/60 nm-thick) was involved as the metallic gate electrode, and different high-*k* dielectrics were used as the gate insulator of the JoFETs. Specifically, we compared 30 nm-thick $HfO_2$ or $Al_2O_3$ layers grown via Atomic Layer Deposition (ALD). The growth was performed at an optimized temperature of 130 °C to preserve the Al/InAs superconducting and transport properties [45]. At 3 K, the fabricated insulators exhibited a relative permittivity ($\varepsilon_R$) of 16.5 and 7.3, and a dielectric strength ($E_{BD}$) of 370 MV/m and 330 MV/m, for $HfO_2$ and $Al_2O_3$, respectively [46].

We fabricated superconductor-semiconductor-superconductor Al-InAs-Al JoFETs with width ($W_{JJ}$) of 6 µm, interelectrode separation ($L_{JJ}$) ranging from 500 to 1250 nm, and gate lengths ($L_G$) of 200 and 1000 nm. Figure 1b depicts an optical microscope photograph of the overall JoFET showing the 4-terminals architecture to which the electrical characterization was referred. Figure 1c shows the Scanning Electron Microscopy (SEM) micrograph of a JoFET with $L_{JJ}$ = 800 nm and $L_G$ = 200 nm, exhibiting the alignment between the metallic gate and the Al/InAs Josephson Junction (JJ).

JoFETs were measured in a dilution cryostat equipped with a z-axis superconducting magnet and a DC measurement setup; the electrical characterization was performed at 50 mK unless stated otherwise. Figure 2a compares gate-dependent V-I curves of JoFETs with $HfO_2$ ($HfO_2$-JoFET) and $Al_2O_3$ ($Al_2O_3$-JoFET) as gate insulators, $W_{JJ}$ = 6 µm, $L_{JJ}$ = 500 nm, and $L_G$= 1000 nm, which are symmetric for positive and negative $I_{DS}$ values. The JoFETs exhibit a non-dissipative behavior (P=0 W) for $I_{DS}$ lower than the switching current ($I_S$), i.e., the maximum non-dissipative current supported. For $I_{DS} > I_S$, the JoFETs feature a dissipative behavior with a power dissipation of $P = V_{DS} \times I_{DS} = R_N \times I_{DS}^2$, where $R_N$ is the normal state resistance, the resistance value calculated from the V-I curve slope for $I>I_S$. A reduction of $I_S$, together with an increase of the $R_N$, was observed as due to the increase of the absolute negative gate voltage, which is related to the depletion of the InAs epilayer from electrons.  This is confirmed for all the JoFETs fabricated, regardless of the different geometries ($L_{JJ}$ and $L_G$) and gate insulators chosen (Figure S1).

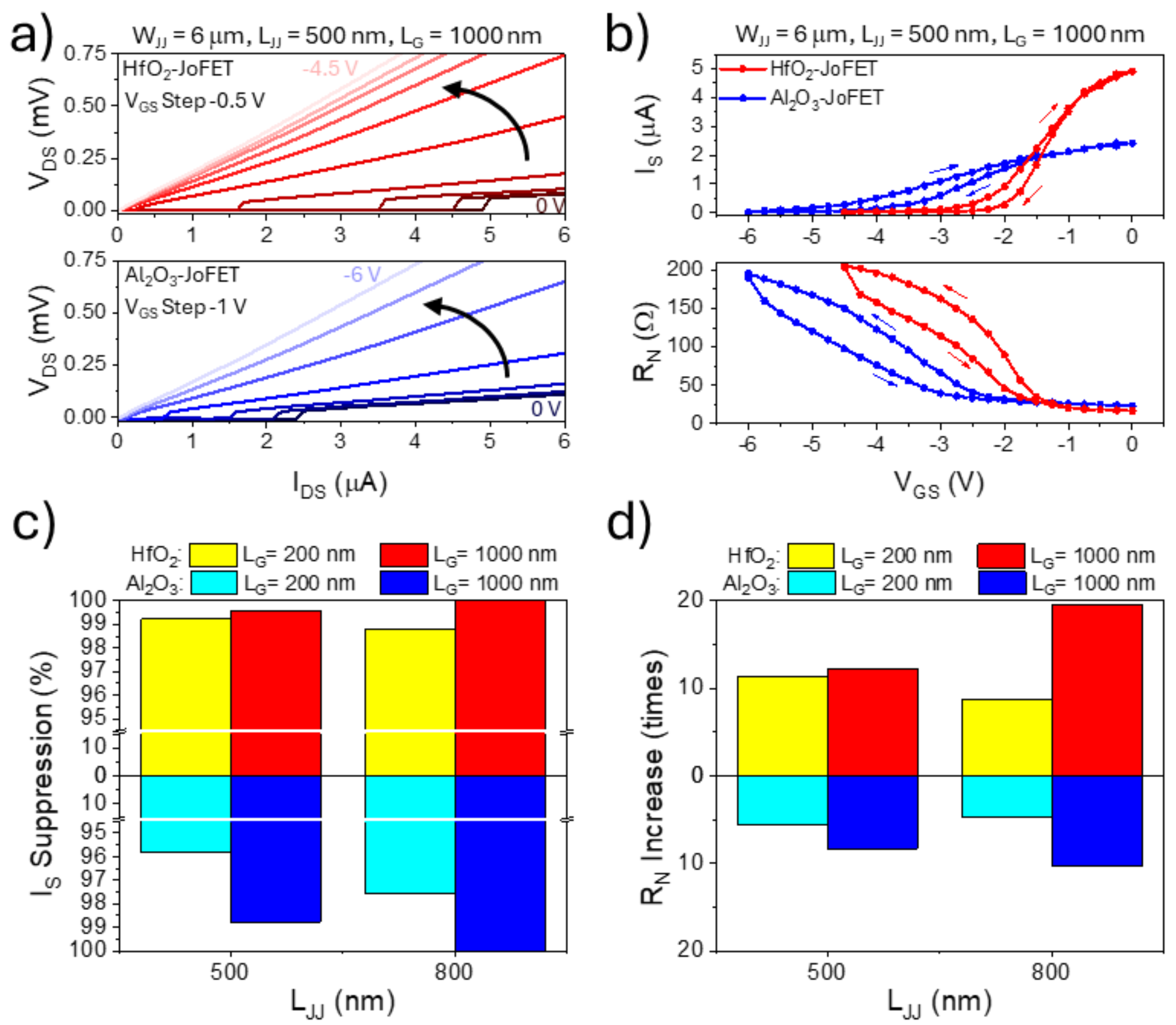

**Figure 2: Gate insulator-dependent electrical behavior of InAsOI-based Josephson Field Effect Transistors.** a) Gate-dependent voltage vs. current characteristics of an $HfO_2$- (up) and an $Al_2O_3$- (bottom) JoFET recorded during the $V_{GS}$ downward scan. b) Upward and downward gate-dependent switching current and normal-state resistance of a $HfO_2$- (up) and an $Al_2O_3$- (bottom) JoFET. The JoFETs feature $W_{JJ}$ = 6 μm, $L_{JJ}$ = 500 nm, and $L_G$= 1000 nm. c,d) Switching current suppression (c) and normal state resistance increase (d) of $HfO_2$- and $Al_2O_3$- JoFETs with different $L_{JJ}$ and $L_G$ values. The measurement temperature is 50 mK.

$I_S$ vs $V_{GS}$ (Figure 2b top) and $R_N$ vs $V_{GS}$ (Figure 2b bottom) trends of $HfO_2$- and $Al_2O_3$-JoFETs extrapolated from Figure 2a are shown in Figure 2b. For both the JoFETs, sigmoidal behaviors of the collected trends were observed: the higher the negative gate voltage absolute value, the lower (higher) the switching current (normal-state resistance), and vice versa. Regardless of the insulator chosen, a hysteresis appears between upward and downward scans, which is likely to be related to charge traps at the InAs/insulator interface since a negligible hysteresis was observed in metal-insulator-metal capacitors featuring the same dielectrics [46]. During the upward $V_{GS}$ scan, a higher quantity of electrons contributes to the electrical transport; consequently, the JoFET features higher $I_S$ and lower $R_N$. On the other hand, during the downward $V_{GS}$ scan, lower electrons are free to contribute to the electrical transport, so that the JoFET exhibits reduced $I_S$ and increased $R_N$. Different gate voltage ranges were chosen for different insulators to reach the plateau of the sigmoidal trend, namely, from 0 V to -4.5 V for $HfO_2$-JoFETs and from 0 V to -6 V for $Al_2O_3$-

JoFETs. The reduced range used for the $HfO_2$-JoFETs is enabled by the higher permittivity exhibited by the insulator [46]. In both cases, the applied gate voltages ensure the use of gate electric fields lower than the dielectric strengths measured for the deposited insulators (Figure S2). This avoids hot quasi-particle injection assisted by the avalanche dielectric breakdown to suppress the switching current value. The specific zero-gate-voltage and gate-dependent electrical properties of JoFETs rely on both the transistor morphology and gate insulator chosen. As expected, by increasing $L_{JJ}$ at $V_{GS}$ = 0 V, we observed a decrease in $I_S$ and an increase in $R_N$ (Figure S3). This is consistent with the zero-gate-voltage switching current density ($I_S/W_{JJ}$) vs $L_{JJ}$ behavior shown in Figure S4, where $I_S/W$ exponentially decreases with the interelectrode separation. When $V_{GS}$=0V was applied to the JoFETs, $HfO_2$-JoFETs exhibited a switching current density ~2 times higher than that achieved with $Al_2O_3$. At the same time, $HfO_2$-JoFETs also feature a normal-state resistance reduced by 30 ÷ 40 % compared to that of $Al_2O_3$-JoFETs. This is related to a lower sheet carrier density of $Al_2O_3$-JoFETs compared to $HfO_2$-JoFETs. In both cases, JoFETs exhibited a switching current density larger than that achieved with bare ungated JJs (Figure S4). Similarly, the impact of the insulator on the zero-gate-voltage electrical performance of JoFETs was also previously reported in superconducting Al/InAs quantum wells (QWs) [31,45]. Following the observations, this could be related to charge trapping mediated by specific positive charged defects owned by each insulator at the dielectric/InAs interface [31,47]. Interestingly, the higher the gate insulator permittivity (16.5 for $HfO_2$, instead of 7.3 and 1 for $Al_2O_3$ and vacuum, respectively), the higher the switching current density featured by the JoFETs. However, we do not propose any phenomenological model for this. The current observation suggests that the choice of gate insulator significantly affects the zero-gate-voltage electrical performance of the fabricated JoFETs. This behavior was consistently observed across all JoFETs in this study, regardless of their morphological properties (Figure S1).

We evaluated the electrical performance of JoFETs by using two figures of merits (FOMs), namely, the $I_S$ *suppression* factor ($I_{Sup} = 1 - \frac{I_{S@V_{GS_{min}}}}{I_{S@V_{GS}=0V}}$) and the $R_N$ *increase* factor ($R_{Inc} = \frac{R_{N@V_{GS_{min}}}}{R_{N@V_{GS}=0V}}$) (Figures 2c,d) [26]. Depending on the gate insulator chosen, the FOMs were evaluated by using the minimum value of $V_{GS}$ applied ($V_{GS_{min}}$), namely, -4.5 V for $HfO_2$-JoFETs and -6 V for $Al_2O_3$-JoFETs. For both the gate insulators and interelectrode separations tested ($L_{JJ}$ = 500 or 800 nm), higher $I_{Sup}$ and $R_{Inc}$ factors were obtained in the case of wider gates compared to the shorter ones. Moreover, longer JJs and wider gates yield the best $I_{Sup}$ and $R_{Inc}$ values for both insulators. In the case of the $HfO_2$-JoFET with $L_G$=1000 nm and $L_{JJ}$=800 nm, a 100 % switching current suppression and a ~20-times increase of the normal-state resistance were obtained by changing the gate voltage in the range from 0 to -4.5 V (Figure S1f). On the other hand, the $Al_2O_3$-JoFET counterpart exhibits only ~10 times the increase of the normal-state resistance, although the total suppression of the switching current was achieved (Figure S1f). Since both the JoFETs reached the same plateau value of ~550 Ω at $V_{GS_{min}}$, as reported in Figure S1f, the $R_{Inc}$ factors agree if we take into account the higher value of $R_N$ at 0-gate-voltage of the $AlO_3$-JoFET (50 Ω) compared to the $HfO_2$-JoFET (30 Ω).

We then evaluated the temperature-dependent behavior of the fabricated JoFETs. Figures 3a and b show $I_S$ vs $V_{GS}$ and $R_N$ vs $V_{GS}$ trends of a $HfO_2$-JoFET (Figure 3a) and an $Al_2O_3$-JoFET (Figure 3b) with $W_{JJ}$ = 6 μm, $L_{JJ}$=500 nm, and $L_G$=1000 nm, in the temperature range from 50 mK to 1000 mK.

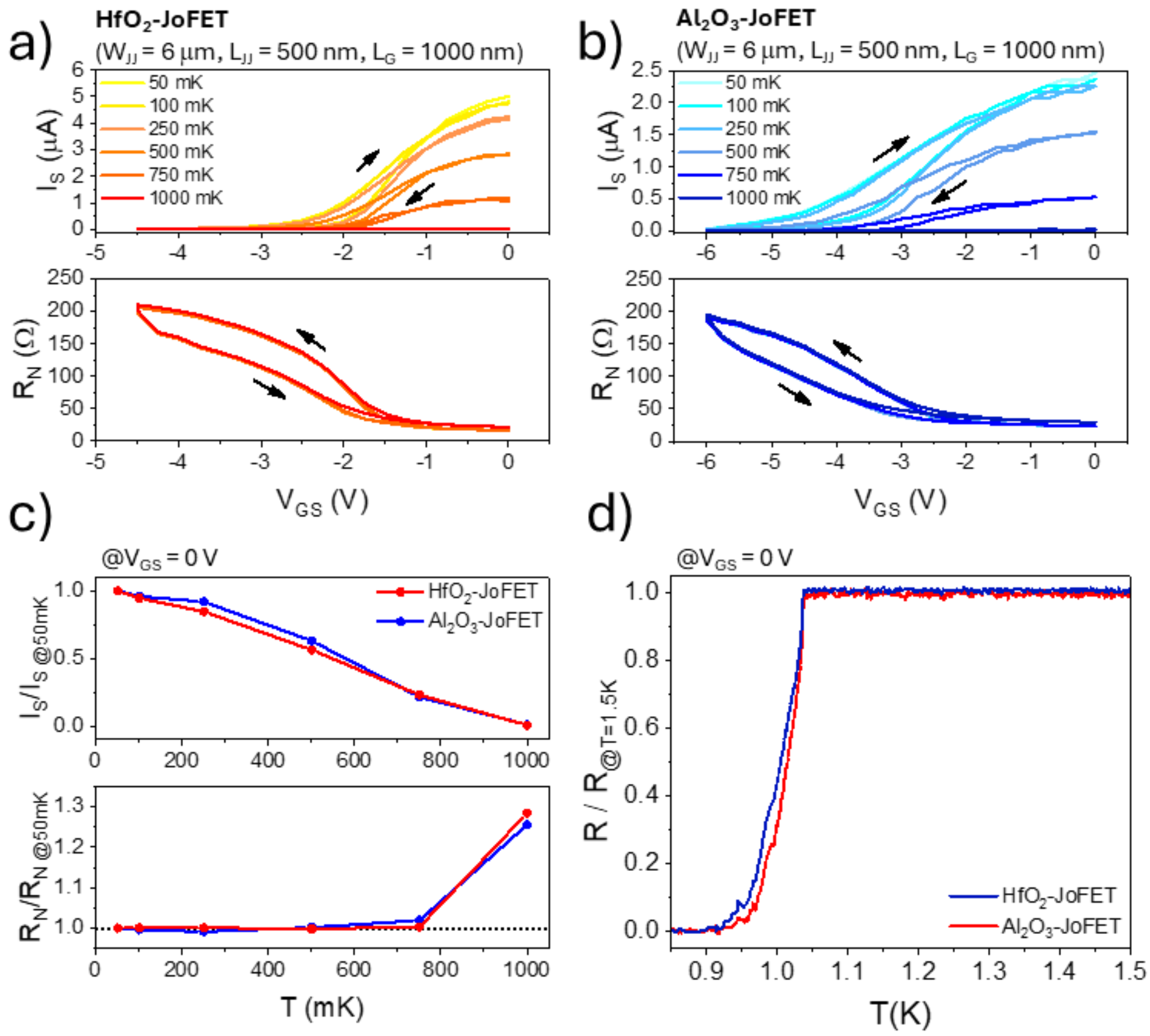

**Figure 3: Temperature-dependent behavior of InAsOI-based Josephson Field Effect Transistors.** a,b) Temperature-dependent behavior of forward and backward gate-dependent switching current and normal-state resistance scans of a $HfO_2$- (a) or $Al_2O_3$- (b) JoFET. Normal-state resistance curves for temperatures lower than 750 mK are indistinguishable and superimposed. c) Temperature-dependent behavior of the zero-gate-voltage switching current (up) and normal-state resistance (bottom) for $HfO_2$- and $Al_2O_3$-JoFETs normalized with respect to the value at 50 mK. d) Temperature-dependent behavior of the 0-gate-voltage resistance of $HfO_2$- and $Al_2O_3$-JoFETs normalized respect to the value at 1.5 K. The JoFETs feature $W_{JJ}$ = 6 μm, $L_{JJ}$ = 500 nm, and $L_G$= 1000 nm.

As expected, the switching current monotonically decreases with the increase in temperature, regardless of the gate voltage and the gate insulator chosen, until it is flattened to 0 A at 1 K. At $V_{GS}$ = 0 V, the $\frac{I_S}{I_{S@50mK}}$ ratio reduces in agreement with what is also observed for InAsOI-based JJs (Figure 3c top) [12]. Differently, the $R_N$ vs. $V_{GS}$ trend remains unchanged until 1 K, which is also immediately apparent from Figure 3c bottom, where the $\frac{R_N}{R_{N@50mK}}$ ratio at $V_{GS}$ = 0 V was reported. The latter result agrees with the resistance vs. temperature behavior shown in Figure 3d. The measured resistance, which is the series of the Al leads and the InAs junction, gets higher due to the Al transition to normal metal. Tested JoFETs show a critical temperature ($T_C$) of ~1 K, comparable to that achieved with bare JJs without gate architectures [12]. It is worth noting that the $T_C$ of the JoFETs is also equal to that of the superconducting Al since no other transition in the resistance is visible until 1.5 K. We report that, despite the $T_C$ of 1 K achieved for Al on this experimental run, Al films (with and without oxide deposited via ALD) produced in a successive run feature a $T_C$ of 1.2 K (Figure S5). This experiment infers that the deposition of an insulator through ALD at the temperature of 130 °C for tens of hours does not change the superconducting properties of an Al film of 100 nm in thickness. Despite the best we did during the manufacturing process, we conclude that the Al film we used for the JoFETs exhibited a reduced $T_C$ of 1 K.

We next collected JoFETs' behavior under different values of the out-of-plane magnetic field ($B_\perp$) in the range ±1 mT. Figures 4a and b show the gate-dependent switching current vs. $B_\perp$ trend for an $HfO_2$-JoFET (Figure 4a) and an $Al_2O_3$-JoFET (Figure 4b) with $W_{JJ}$=6 μm, $L_{JJ}$=500 nm, and $L_G$=1000 nm. $I_S$ vs. $B_\perp$ characteristics for different superconducting transistors are reported in Figure S6.

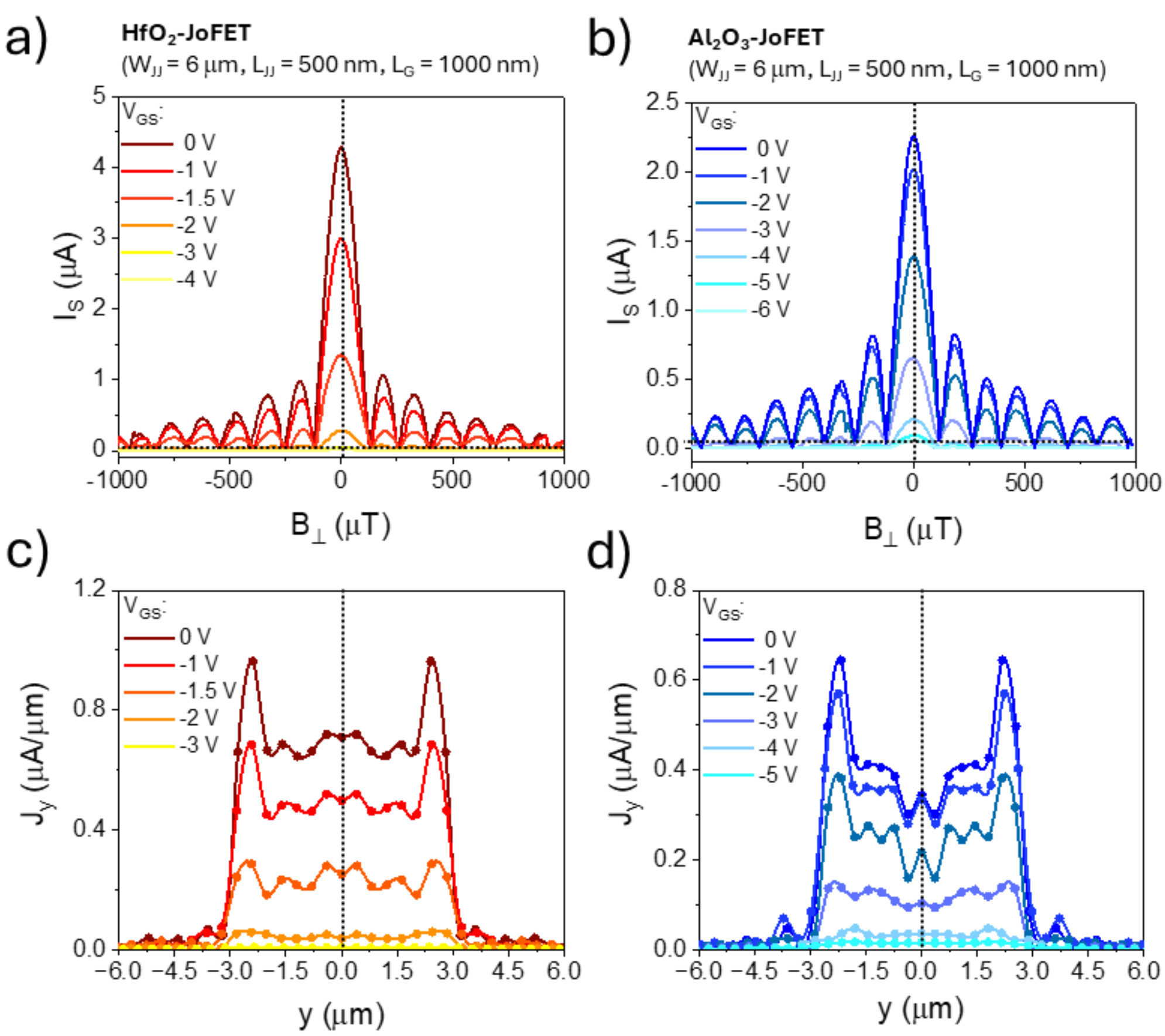

**Figure 4: Out-of-plane magnetic field-dependent behavior of InAsOI-based Josephson Field Effect Transistors.** a,b) Gate-dependent switching current vs. out-of-plane magnetic field characteristic of a $HfO_2$- (a) or $Al_2O_3$- (b) JoFET. c,d) Supercurrent current density distribution vs. width position characteristic of a $HfO_2$- (c) or $Al_2O_3$- (d) JoFET. The JoFETs feature $W_{JJ}$ = 6 µm, $L_{JJ}$ = 500 nm, and $L_G$= 1000 nm. The measurement temperature is 50 mK.

Regardless of the morphology and gate insulator chosen, by applying $B_\perp$ to the JoFET, the switching current roughly follows a Fraunhofer-like pattern $I_S(B_\perp) = I_S|_{B_\perp=0T} \times \left| \frac{sin\left(\frac{\pi B_\perp A_{eff}}{\Phi_0}\right)}{\frac{\pi B_\perp A_{eff}}{\Phi_0}} \right|$, where $A_{eff} = \gamma \times L_{JJ} \times W_{JJ}$ is the JJ effective area and $\Phi_0 = \frac{h}{2e}$ is the flux quantum. The scaling factor $\gamma$ accounts for flux enhancement due to both magnetic field focusing and an increased JJ length compared to the inter-electrode separation. $\gamma$ depends on the morphology exhibiting a monotonic increase with the JJ aspect ratio $\frac{W_{JJ}}{L_{JJ}}$ (Table S1). Values of $\gamma$ are perfectly in agreement with the previously calculated results, indicating no significant influence from the thin superconducting gate electrode [12]. Due to the weak paramagnetism of the insulators chosen, a negligible impact of the insulator type and thickness is expected in flux focusing and supercurrent distribution. The magnitude of the interference pattern decreases at lower gate voltage values, while the pattern periodicity, related to $A_{eff}$, remains largely unchanged across the tested gate voltages.

Deviations from the conventional Fraunhofer-like diffraction pattern are evident in the reduced damping of secondary peaks and the non-ideal periodicity of the zeros.

We then computed the inverse Fast Fourier Transform (iFFT) of the collected $I_S$ vs. $B_\perp$ patterns to obtain the supercurrent density distribution ($J_y$) along the width of the JJ. $J_y$ was calculated by assuming an ideal sinusoidal current-phase relation, while a deviation from this approximation does not provide a better description. Figures 4c and d show the $J_y$ vs. width position evaluated from the interference patterns shown in Figures 4a and b. Instead of the expected symmetric squared iFFT, evident fluctuations in the supercurrent density are observed with significant peaks at the mesa edges $|y| \approx \frac{W_{JJ}}{2}$, indicating a more profound deviation of the diffraction pattern with respect to the Fraunhofer approximation. Deviations from Fraunhofer to Gaussian diffraction patterns are possible for narrow JJs [48,49], i.e., $W_{JJ} < \xi_0$, where $\xi_0$ is the superconducting coherence length in the proximitized InAs JJ. With a typical value of $\xi_0 \approx 650$ nm, this hypothesis is excluded.

The deviation from the Fraunhofer approximation can be explained by (i) a discretization artifact due to the finite number of measured periods in the interference pattern, (ii) an increased carrier density at the edge of the mesa due to band bending of the lateral confinement potential, (iii) a nonuniform distribution of magnetic fields in the JJ due to flux focusing from the superconducting leads.

To test (i), we computed the iFFT of the absolute value of an ideal sinc-function employing 0-gate-voltage $I_S|_{B_\perp=0T}$ and $W_{JJ}$ equal to that of the $HfO_2$-JoFET shown in Figure 4a. The sinc-function was truncated at the same number of periods of the measured Fraunhofer pattern, namely 15 $\Phi_0$. Figure S7 shows the comparison of the calculated supercurrent density distributions. The iFFT of the truncated sinc-function shows supercurrent density peaks at the edges ~5 times smaller than those obtained for the measured Fraunhofer pattern, ruling out the low resolution in the y-axis position as an explanation of the observed peaks. On the other hand, the oscillating behavior observed in $J_y$ inside the mesa can be ascribed to the finite number of lobes in the Fraunhofer pattern, as the same features are retrieved when computing $J_y$ from the theoretical sinc-function.

An increased supercurrent density at the edges of the JJ is typically associated with the InAs band bending at the mesa edges (ii), leading to higher electron density compared to the rest of the InAs mesa [50,51]. This hypothesis is ruled out by considering that anomalies in the interference diffraction pattern were not observed for InAsOI-based JJs featuring superconducting leads reduced in length to 500 nm [25], instead of 100 μm as reported in this work (Figure S8). Moreover, the typical extension of the surface accumulation layer is on the order of tens of nanometers [50]. It does not fully account for the hundreds-of-nanometers width of the peaks observed in most of the iFFT (Figure 4 and S6).

We propose a non-uniform flux focusing (iii) at the origin of such an anomaly. In planar JJ geometry, Meissner screening tends to concentrate the field lines in the middle of the JJ from the Al superconducting leads, thereby distorting the traditional Fraunhofer-like pattern [52–54]. The distortion in the diffraction pattern depends on the geometry of the leads and the aspect ratio of the JJ ($\frac{W_{JJ}}{L_{JJ}}$). Non-uniform magnetic field distribution becomes significant when $\frac{W_{JJ}}{L_{JJ}} \geq 1$, reducing the damping of the interference pattern from the usual $1/B$ to $\sim 1/\sqrt{B}$ for $\frac{W_{JJ}}{L_{JJ}} \gg 1$ [52]. In this asymptotic limit, the effective junction area scales as $A_{eff} \sim 1.842$, then increasing the periodicity in $B_{\perp}$ and the zeroes of $I_S(B_{\perp})$ are not fully periodic in $B_{\perp}$ [54]. These features are consistent with the measured diffraction patterns, which show a non-ideal periodicity of the zeroes and a lower damping of the side lobes compared to a conventional Fraunhofer-like pattern.

By reducing the gate voltage, the magnitude of the side peaks in the iFFT tends to decrease. This effect can be ascribed to a non-uniform field effect. Indeed, when the gate potential is applied to deplete the InAs transport channels, the electrical field lines have a higher density at the edges of the mesa structure side covered by the metallic gate. This results in a higher electrical field at the edges of the mesa and more efficient suppression of the edge electron density, decreasing the impact of the edge effect in the diffraction pattern for both hypotheses (ii) and (iii). Similar results are observed for all the JoFETs fabricated in this work with extended leads, regardless of the morphological properties ($L_{JJ}$ and $L_G$) and gate insulator chosen (Figure S6).

**Conclusions**

In summary, we studied the gate-tunable electrical properties of InAsOI-based JoFETs by employing $HfO_2$ or $Al_2O_3$ as gate insulators. For both the high-*k* dielectrics, a total suppression of the switching current and 10-20 times increase of the normal state resistance were obtained with negative gate voltages. Thanks to the higher permittivity of the insulator, $HfO_2$-JoFETs are more efficient in tuning the switching current and the normal state resistance values with the gate voltage, compared to $Al_2O_3$-JoFETs. Temperature (50 – 1000 mK) and out-of-plane magnetic field (-1 – 1 mT) dependent behaviors of InAsOI-based JoFETs were evaluated. We observed a decrease in the switching current with the temperature increase, regardless of the gate voltage and the gate insulator chosen. On the other hand, the normal state resistance vs. gate voltage trend remains unchanged until 1 K. With the out-of-plane magnetic field, JoFETs exhibited diffraction patterns distorted with respect to the conventional Fraunhofer-like interference. The origin of such anomalies can be identified in the physics of the JJ edges, either with an increased current density or with a more accurate consideration of non-uniform flux focusing on the superconducting leads. The results achieved represent an important insight into the InAsOI-based JoFET electrical behavior.

**Supporting Information**

The supporting information document contains supplementary Figures S1-S9 and includes materials, manufacturing methods, and characterization methods. The supercurrent density estimation was detailed reported as well as the Python script used. The estimation of the coherence length and the electron mean free was described.

**Acknowledgments**

This work was partially supported by H2020-EU.1.2. - EXCELLENT SCIENCE - Future and Emerging Technologies (FET) under grant 964398 (SuperGate), by HORIZON.3.1 - The European Innovation Council (EIC) – Transition Open programme under grant 101057977 (SPECTRUM), and by the Piano Nazionale di Ripresa e Resilienza, Ministero dell'Università e della Ricerca (PNRR MUR) Project under Grant PE0000023-NQSTI.

**References**

(1) Chen, O.; Cai, R.; Wang, Y.; Ke, F.; Yamae, T.; Saito, R.; Takeuchi, N.; Yoshikawa, N. Adiabatic Quantum-Flux-Parametron: Towards Building Extremely Energy-Efficient Circuits and Systems. *Sci. Rep.* **2019**, *9* (1), 10514. https://doi.org/10.1038/s41598-019-46595-w.

(2) Likharev, K. K.; Semenov, V. K. RSFQ Logic/Memory Family: A New Josephson-Junction Technology for Sub-Terahertz-Clock-Frequency Digital Systems. *IEEE Trans. Appl. Supercond.* **1991**, *1* (1), 3–28. https://doi.org/10.1109/77.80745.

(3) De Franceschi, S.; Giazotto, F.; Beltram, F.; Sorba, L.; Lazzarino, M.; Franciosi, A. Andreev Reflection in Si-Engineered Al/InGaAs Hybrid Junctions. *Appl. Phys. Lett.* **1998**, *73* (26), 3890–3892. https://doi.org/10.1063/1.122926.

(4) Giazotto, F.; Pingue, P.; Beltram, F.; Lazzarino, M.; Orani, D.; Rubini, S.; Franciosi, A. Resonant Transport in Nb/GaAs/AlGaAs Heterostructures: Realization of the de Gennes–Saint-James Model. *Phys. Rev. Lett.* **2001**, *87* (21), 216808. https://doi.org/10.1103/PhysRevLett.87.216808.

(5) Roddaro, S.; Pescaglini, A.; Ercolani, D.; Sorba, L.; Giazotto, F.; Beltram, F. Hot-Electron Effects in InAs Nanowire Josephson Junctions. *Nano Res.* **2011**, *4* (3), 259–265. https://doi.org/10.1007/s12274-010-0077-6.

(6) Tiira, J.; Strambini, E.; Amado, M.; Roddaro, S.; San-Jose, P.; Aguado, R.; Bergeret, F. S.; Ercolani, D.; Sorba, L.; Giazotto, F. Magnetically-Driven Colossal Supercurrent Enhancement in InAs Nanowire Josephson Junctions. *Nat. Commun.* **2017**, *8* (1), 14984. https://doi.org/10.1038/ncomms14984.

(7) Scappucci, G.; Kloeffel, C.; Zwanenburg, F. A.; Loss, D.; Myronov, M.; Zhang, J.-J.; De Franceschi, S.; Katsaros, G.; Veldhorst, M. The Germanium Quantum Information Route. *Nat. Rev. Mater.* **2020**, *6* (10), 926–943. https://doi.org/10.1038/s41578-020-00262-z.

(8) Burkard, G.; Gullans, M. J.; Mi, X.; Petta, J. R. Superconductor–Semiconductor Hybrid-Circuit Quantum Electrodynamics. *Nat. Rev. Phys.* **2020**, *2* (3), 129–140. https://doi.org/10.1038/s42254-019-0135-2.

(9) Akazaki, T.; Takayanagi, H.; Nitta, J.; Enoki, T. A Josephson Field Effect Transistor Using an InAs-Inserted-Channel In0.52Al0.48As/In0.53Ga0.47As Inverted Modulation-Doped Structure. *Appl. Phys. Lett.* **1996**, *68* (3), 418–420. https://doi.org/10.1063/1.116704.

(10) Doh, Y.-J.; van Dam, J. A.; Roest, A. L.; Bakkers, E. P. A. M.; Kouwenhoven, L. P.; De Franceschi, S. Tunable Supercurrent Through Semiconductor Nanowires. *Science (80-. ).*

**2005**, *309* (5732), 272–275. https://doi.org/10.1126/science.1113523.

(11) Shabani, J.; Kjaergaard, M.; Suominen, H. J.; Kim, Y.; Nichele, F.; Pakrouski, K.; Stankevic, T.; Lutchyn, R. M.; Krogstrup, P.; Feidenhans'l, R.; Kraemer, S.; Nayak, C.; Troyer, M.; Marcus, C. M.; Palmstrøm, C. J. Two-Dimensional Epitaxial Superconductor-Semiconductor Heterostructures: A Platform for Topological Superconducting Networks. *Phys. Rev. B* **2016**, *93* (15), 155402. https://doi.org/10.1103/PhysRevB.93.155402.

(12) Paghi, A.; Trupiano, G.; De Simoni, G.; Arif, O.; Sorba, L.; Giazotto, F. InAs on Insulator: A New Platform for Cryogenic Hybrid Superconducting Electronics. *Adv. Funct. Mater.* **2025**, *35* (7). https://doi.org/10.1002/adfm.202416957.

(13) Delfanazari, K.; Puddy, R. K.; Ma, P.; Yi, T.; Cao, M.; Gul, Y.; Farrer, I.; Ritchie, D. A.; Joyce, H. J.; Kelly, M. J.; Smith, C. G. On-Chip Andreev Devices: Hard Superconducting Gap and Quantum Transport in Ballistic Nb–In 0.75 Ga 0.25 As-Quantum-Well–Nb Josephson Junctions. *Adv. Mater.* **2017**, *29* (37), 0–5. https://doi.org/10.1002/adma.201701836.

(14) Delfanazari, K.; Li, J.; Xiong, Y.; Ma, P.; Puddy, R. K.; Yi, T.; Farrer, I.; Komori, S.; Robinson, J. W. A.; Serra, L.; Ritchie, D. A.; Kelly, M. J.; Joyce, H. J.; Smith, C. G. Quantized Conductance in Hybrid Split-Gate Arrays of Superconducting Quantum Point Contacts with Semiconducting Two-Dimensional Electron Systems. *Phys. Rev. Appl.* **2024**, *21* (1), 014051. https://doi.org/10.1103/PhysRevApplied.21.014051.

(15) Delfanazari, K.; Li, J.; Ma, P.; Puddy, R. K.; Yi, T.; Xiong, Y.; Farrer, I.; Komori, S.; Robinson, J. W. A.; Ritchie, D. A.; Kelly, M. J.; Joyce, H. J.; Smith, C. G. Large-Scale On-Chip Integration of Gate-Voltage Addressable Hybrid Superconductor–Semiconductor Quantum Wells Field Effect Nano-Switch Arrays. *Adv. Electron. Mater.* **2024**, *10* (2), 1–9. https://doi.org/10.1002/aelm.202300453.

(16) Marsh, A. M.; Williams, D. A.; Ahmed, H. Supercurrent Transport through a High-Mobility Two-Dimensional Electron Gas. *Phys. Rev. B* **1994**, *50* (11), 8118–8121. https://doi.org/10.1103/PhysRevB.50.8118.

(17) Wan, Z.; Kazakov, A.; Manfra, M. J.; Pfeiffer, L. N.; West, K. W.; Rokhinson, L. P. Induced Superconductivity in High-Mobility Two-Dimensional Electron Gas in Gallium Arsenide Heterostructures. *Nat. Commun.* **2015**, *6* (1), 7426. https://doi.org/10.1038/ncomms8426.

(18) Meissner, H. Superconductivity of Contacts with Interposed Barriers. *Phys. Rev.* **1960**, *117* (3), 672–680. https://doi.org/10.1103/PhysRev.117.672.

(19) Pannetier, B.; Courtois, H. Andreev Reflection and Proximity Effect. *J. Low Temp. Phys.*

**2000**, *118*, 599–615. https://doi.org/10.1023/A:1004635226825.

(20) Casparis, L.; Connolly, M. R.; Kjaergaard, M.; Pearson, N. J.; Kringhøj, A.; Larsen, T. W.; Kuemmeth, F.; Wang, T.; Thomas, C.; Gronin, S.; Gardner, G. C.; Manfra, M. J.; Marcus, C. M.; Petersson, K. D. Superconducting Gatemon Qubit Based on a Proximitized Two-Dimensional Electron Gas. *Nat. Nanotechnol.* **2018**, *13* (10), 915–919. https://doi.org/10.1038/s41565-018-0207-y.

(21) Larsen, T. W.; Petersson, K. D.; Kuemmeth, F.; Jespersen, T. S.; Krogstrup, P.; Nygård, J.; Marcus, C. M. Semiconductor-Nanowire-Based Superconducting Qubit. *Phys. Rev. Lett.* **2015**, *115* (12), 127001. https://doi.org/10.1103/PhysRevLett.115.127001.

(22) Paajaste, J.; Amado, M.; Roddaro, S.; Bergeret, F. S.; Ercolani, D.; Sorba, L.; Giazotto, F. Pb/InAs Nanowire Josephson Junction with High Critical Current and Magnetic Flux Focusing. *Nano Lett.* **2015**, *15* (3), 1803–1808. https://doi.org/10.1021/nl504544s.

(23) Drachmann, A. C. C.; Suominen, H. J.; Kjaergaard, M.; Shojaei, B.; Palmstrøm, C. J.; Marcus, C. M.; Nichele, F. Proximity Effect Transfer from NbTi into a Semiconductor Heterostructure via Epitaxial Aluminum. *Nano Lett.* **2017**, *17* (2), 1200–1203. https://doi.org/10.1021/acs.nanolett.6b04964.

(24) Telkamp, S.; Antonelli, T.; Todt, C.; Hinderling, M.; Coraiola, M.; Haxell, D.; Kate, S. C. ten; Sabonis, D.; Zeng, P.; Schott, R.; Cheah, E.; Reichl, C.; Nichele, F.; Krizek, F.; Wegscheider, W. Development of a Nb-Based Semiconductor-Superconductor Hybrid 2DEG Platform. *Adv. Electron. Mater.* **2025**, *2400687*, 1–12. https://doi.org/10.1002/aelm.202400687.

(25) Battisti, S.; Simoni, G. De; Braggio, A.; Paghi, A.; Sorba, L.; Giazotto, F. Extremely Weak Sub-Kelvin Electron–Phonon Coupling in InAs on Insulator. *Appl. Phys. Lett.* **2024**, *125* (20). https://doi.org/10.1063/5.0225361.

(26) Paghi, A.; Borgongino, L.; Tortorella, S.; De Simoni, G.; Strambini, E.; Sorba, L.; Giazotto, F. Supercurrent Multiplexing with Solid-State Integrated Hybrid Superconducting Electronics. *Prepr. https//arxiv.org/abs/2410.11721* **2024**.

(27) Haxell, D. Z.; Coraiola, M.; Sabonis, D.; Hinderling, M.; ten Kate, S. C.; Cheah, E.; Krizek, F.; Schott, R.; Wegscheider, W.; Nichele, F. Zeeman- and Orbital-Driven Phase Shifts in Planar Josephson Junctions. *ACS Nano* **2023**, *17* (18), 18139–18147. https://doi.org/10.1021/acsnano.3c04957.

(28) Bøttcher, C. G. L.; Nichele, F.; Shabani, J.; Palmstrøm, C. J.; Marcus, C. M. Dynamical Vortex Transitions in a Gate-Tunable Two-Dimensional Josephson Junction Array. *Phys.*

*Rev. B* **2023**, *108* (13), 134517. https://doi.org/10.1103/PhysRevB.108.134517.

(29) Danilenko, A.; Pöschl, A.; Sabonis, D.; Vlachodimitropoulos, V.; Thomas, C.; Manfra, M. J.; Marcus, C. M. Spin Spectroscopy of a Hybrid Superconducting Nanowire Using Side-Coupled Quantum Dots. *Phys. Rev. B* **2023**, *108* (5), 054514. https://doi.org/10.1103/PhysRevB.108.054514.

(30) Wen, F.; Shabani, J.; Tutuc, E. Josephson Junction Field-Effect Transistors for Boolean Logic Cryogenic Applications. *IEEE Trans. Electron Devices* **2019**, *66* (12), 5367–5374. https://doi.org/10.1109/TED.2019.2951634.

(31) Barati, F.; Thompson, J. P.; Dartiailh, M. C.; Sardashti, K.; Mayer, W.; Yuan, J.; Wickramasinghe, K.; Watanabe, K.; Taniguchi, T.; Churchill, H.; Shabani, J. Tuning Supercurrent in Josephson Field-Effect Transistors Using h-BN Dielectric. *Nano Lett.* **2021**, *21* (5), 1915–1920. https://doi.org/10.1021/acs.nanolett.0c03183.

(32) Phan, D.; Falthansl-Scheinecker, P.; Mishra, U.; Strickland, W. M.; Langone, D.; Shabani, J.; Higginbotham, A. P. Gate-Tunable Superconductor-Semiconductor Parametric Amplifier. *Phys. Rev. Appl.* **2023**, *19* (6), 064032. https://doi.org/10.1103/PhysRevApplied.19.064032.

(33) Ciaccia, C.; Haller, R.; Drachmann, A. C. C.; Lindemann, T.; Manfra, M. J.; Schrade, C.; Schönenberger, C. Gate-Tunable Josephson Diode in Proximitized InAs Supercurrent Interferometers. *Phys. Rev. Res.* **2023**, *5* (3), 033131. https://doi.org/10.1103/PhysRevResearch.5.033131.

(34) Casparis, L.; Larsen, T. W.; Olsen, M. S.; Kuemmeth, F.; Krogstrup, P.; Nygård, J.; Petersson, K. D.; Marcus, C. M. Gatemon Benchmarking and Two-Qubit Operations. *Phys. Rev. Lett.* **2016**, *116* (15), 150505. https://doi.org/10.1103/PhysRevLett.116.150505.

(35) Strickland, W. M.; Baker, L. J.; Lee, J.; Dindial, K.; Elfeky, B. H.; Strohbeen, P. J.; Hatefipour, M.; Yu, P.; Levy, I.; Issokson, J.; Manucharyan, V. E.; Shabani, J. Characterizing Losses in InAs Two-Dimensional Electron Gas-Based Gatemon Qubits. *Phys. Rev. Res.* **2024**, *6* (2), 023094. https://doi.org/10.1103/PhysRevResearch.6.023094.

(36) Zheng, H.; Cheung, L. Y.; Sangwan, N.; Kononov, A.; Haller, R.; Ridderbos, J.; Ciaccia, C.; Ungerer, J. H.; Li, A.; Bakkers, E. P. A. M.; Baumgartner, A.; Schönenberger, C. Coherent Control of a Few-Channel Hole Type Gatemon Qubit. *Nano Lett.* **2024**, *24* (24), 7173–7179. https://doi.org/10.1021/acs.nanolett.4c00770.

(37) Strambini, E.; Iorio, A.; Durante, O.; Citro, R.; Sanz-Fernández, C.; Guarcello, C.; Tokatly, I. V.; Braggio, A.; Rocci, M.; Ligato, N.; Zannier, V.; Sorba, L.; Bergeret, F. S.; Giazotto, F. A Josephson Phase Battery. *Nat. Nanotechnol.* **2020**, *15* (8), 656–660.

https://doi.org/10.1038/s41565-020-0712-7.

(38) Iorio, A.; Rocci, M.; Bours, L.; Carrega, M.; Zannier, V.; Sorba, L.; Roddaro, S.; Giazotto, F.; Strambini, E. Vectorial Control of the Spin–Orbit Interaction in Suspended InAs Nanowires. *Nano Lett.* **2019**, *19* (2), 652–657. https://doi.org/10.1021/acs.nanolett.8b02828.

(39) Bohr, M.; Chau, R.; Ghani, T.; Mistry, K. The High-k Solution. *IEEE Spectr.* **2007**, *44* (10), 29–35. https://doi.org/10.1109/MSPEC.2007.4337663.

(40) Javey, A.; Kim, H.; Brink, M.; Wang, Q.; Ural, A.; Guo, J.; McIntyre, P.; McEuen, P.; Lundstrom, M.; Dai, H. High-κ Dielectrics for Advanced Carbon-Nanotube Transistors and Logic Gates. *Nat. Mater.* **2002**, *1* (4), 241–246. https://doi.org/10.1038/nmat769.

(41) Li, W.; Zhou, J.; Cai, S.; Yu, Z.; Zhang, J.; Fang, N.; Li, T.; Wu, Y.; Chen, T.; Xie, X.; Ma, H.; Yan, K.; Dai, N.; Wu, X.; Zhao, H.; Wang, Z.; He, D.; Pan, L.; Shi, Y.; Wang, P.; Chen, W.; Nagashio, K.; Duan, X.; Wang, X. Uniform and Ultrathin High-κ Gate Dielectrics for Two-Dimensional Electronic Devices. *Nat. Electron.* **2019**, *2* (12), 563–571. https://doi.org/10.1038/s41928-019-0334-y.

(42) Wirths, S.; Weis, K.; Winden, A.; Sladek, K.; Volk, C.; Alagha, S.; Weirich, T. E.; von der Ahe, M.; Hardtdegen, H.; Lüth, H.; Demarina, N.; Grützmacher, D.; Schäpers, T. Effect of Si-Doping on InAs Nanowire Transport and Morphology. *J. Appl. Phys.* **2011**, *110* (5). https://doi.org/10.1063/1.3631026.

(43) Datta, S. *Electronic Transport in Mesoscopic Systems*; Cambridge University Press, 1995. https://doi.org/10.1017/CBO9780511805776.

(44) Yu, P. Y.; Cardona, M. *Fundamentals of Semiconductors*; Graduate Texts in Physics; Springer Berlin Heidelberg: Berlin, Heidelberg, 2010. https://doi.org/10.1007/978-3-642-00710-1.

(45) Sütő, M.; Prok, T.; Makk, P.; Kirti, M.; Biasiol, G.; Csonka, S.; Tóvári, E. Near-Surface InAs Two-Dimensional Electron Gas on a GaAs Substrate: Characterization and Superconducting Proximity Effect. *Phys. Rev. B* **2022**, *106* (23), 235404. https://doi.org/10.1103/PhysRevB.106.235404.

(46) Paghi, A.; Battisti, S.; Tortorella, S.; De Simoni, G.; Giazotto, F. Cryogenic Behavior of High-Permittivity Gate Dielectrics: The Impact of Atomic Layer Deposition Temperature and the Lithographic Patterning Method. *J. Appl. Phys.* **2025**, *137* (4). https://doi.org/10.1063/5.0250428.

(47) Swain, R.; Panda, J.; Jena, K.; Lenka, T. R. Modeling and Simulation of Oxide Dependent 2DEG Sheet Charge Density in AlGaN/GaN MOSHEMT. *J. Comput. Electron.* **2015**, *14* (3),

754–761. https://doi.org/10.1007/s10825-015-0711-3.

(48) Cuevas, J. C.; Bergeret, F. S. Magnetic Interference Patterns and Vortices in Diffusive SNS Junctions. *Phys. Rev. Lett.* **2007**, *99* (21), 217002. https://doi.org/10.1103/PhysRevLett.99.217002.

(49) Bergeret, F. S.; Cuevas, J. C. The Vortex State and Josephson Critical Current of a Diffusive SNS Junction. *J. Low Temp. Phys.* **2008**, *153* (5–6), 304–324. https://doi.org/10.1007/s10909-008-9826-2.

(50) Elfeky, B. H.; Lotfizadeh, N.; Schiela, W. F.; Strickland, W. M.; Dartiailh, M.; Sardashti, K.; Hatefipour, M.; Yu, P.; Pankratova, N.; Lee, H.; Manucharyan, V. E.; Shabani, J. Local Control of Supercurrent Density in Epitaxial Planar Josephson Junctions. *Nano Lett.* **2021**, *21* (19), 8274–8280. https://doi.org/10.1021/acs.nanolett.1c02771.

(51) Wieder, H. H. Surface and Interface Barriers of InxGa1−xAs Binary and Ternary Alloys. *J. Vac. Sci. Technol. B Microelectron. Nanom. Struct. Process. Meas. Phenom.* **2003**, *21* (4), 1915–1919. https://doi.org/10.1116/1.1588646.

(52) Moshe, M.; Kogan, V. G.; Mints, R. G. Edge-Type Josephson Junctions in Narrow Thin-Film Strips. *Phys. Rev. B* **2008**, *78* (2), 020510. https://doi.org/10.1103/PhysRevB.78.020510.

(53) Fermin, R.; de Wit, B.; Aarts, J. Beyond the Effective Length: How to Analyze Magnetic Interference Patterns of Thin-Film Planar Josephson Junctions with Finite Lateral Dimensions. *Phys. Rev. B* **2023**, *107* (6), 064502. https://doi.org/10.1103/PhysRevB.107.064502.

(54) Clem, J. R. Josephson Junctions in Thin and Narrow Rectangular Superconducting Strips. *Phys. Rev. B* **2010**, *81* (14), 144515. https://doi.org/10.1103/PhysRevB.81.144515.

**Supporting Information**

# Josephson Field Effect Transistors with InAs on Insulator and High Permittivity Gate Dielectrics

Alessandro Paghi[1*], Laura Borgongino[1], Sebastiano Battisti[1], Simone Tortorella[1,2], Giacomo Trupiano[1], Giorgio De Simoni[1], Elia Strambini[1], Lucia Sorba[1], and Francesco Giazotto[1]

[1]Istituto Nanoscienze-CNR and Scuola Normale Superiore, Piazza San Silvestro 12, 56127 Pisa, Italy.

[2]Dipartimento di Ingegneria Civile e Industriale, Università di Pisa, Largo Lucio Lazzarino, 56122 Pisa, Italy

[*]Corresponding authors: alessandro.paghi@nano.cnr.it

## Summary

## 1. Supporting Figures

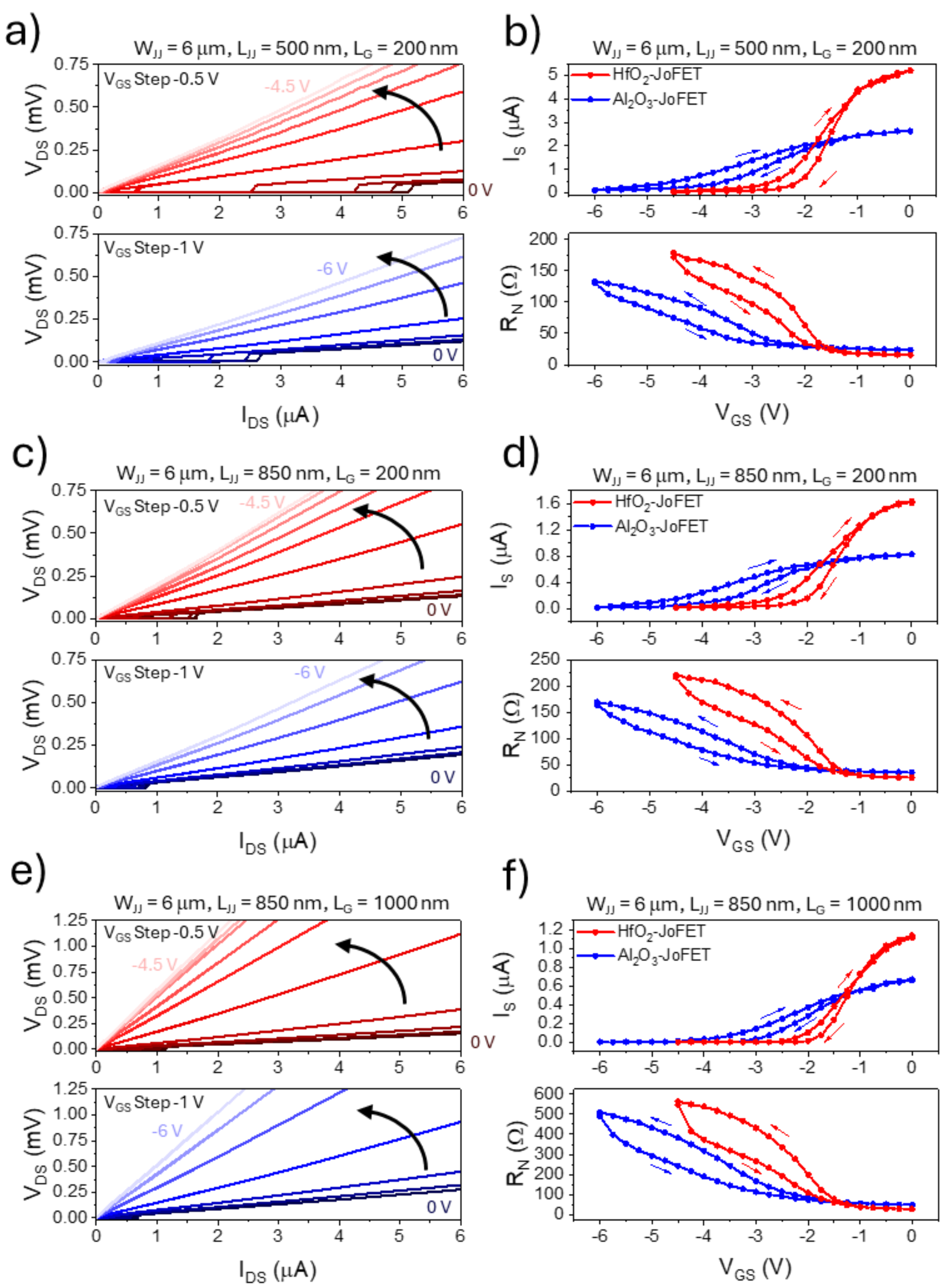

**Figure S1: Electrical characterization of InAsOI-based JoFETs.** a,c,e) Gate-dependent voltage vs. current characteristics of JoFETs. b,d,f) Forward and backward gate-dependent switching current and normal state resistance of JoFETs. Measurements were performed at 50 mK.

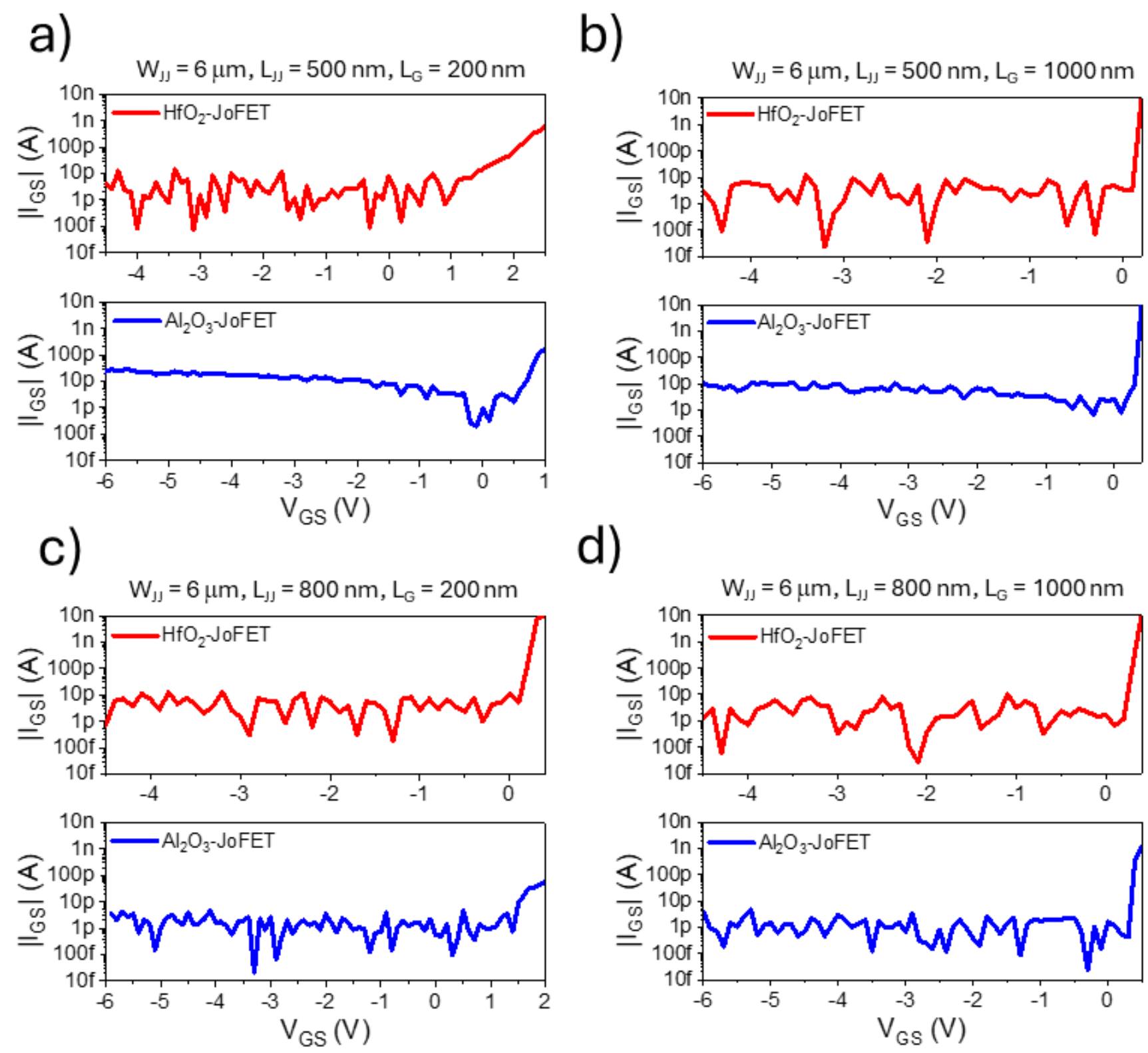

**Figure S2: Gate leakage current vs. gate voltage characteristics of InAsOI-based JoFETs.** Measurements were performed at 50 mK.

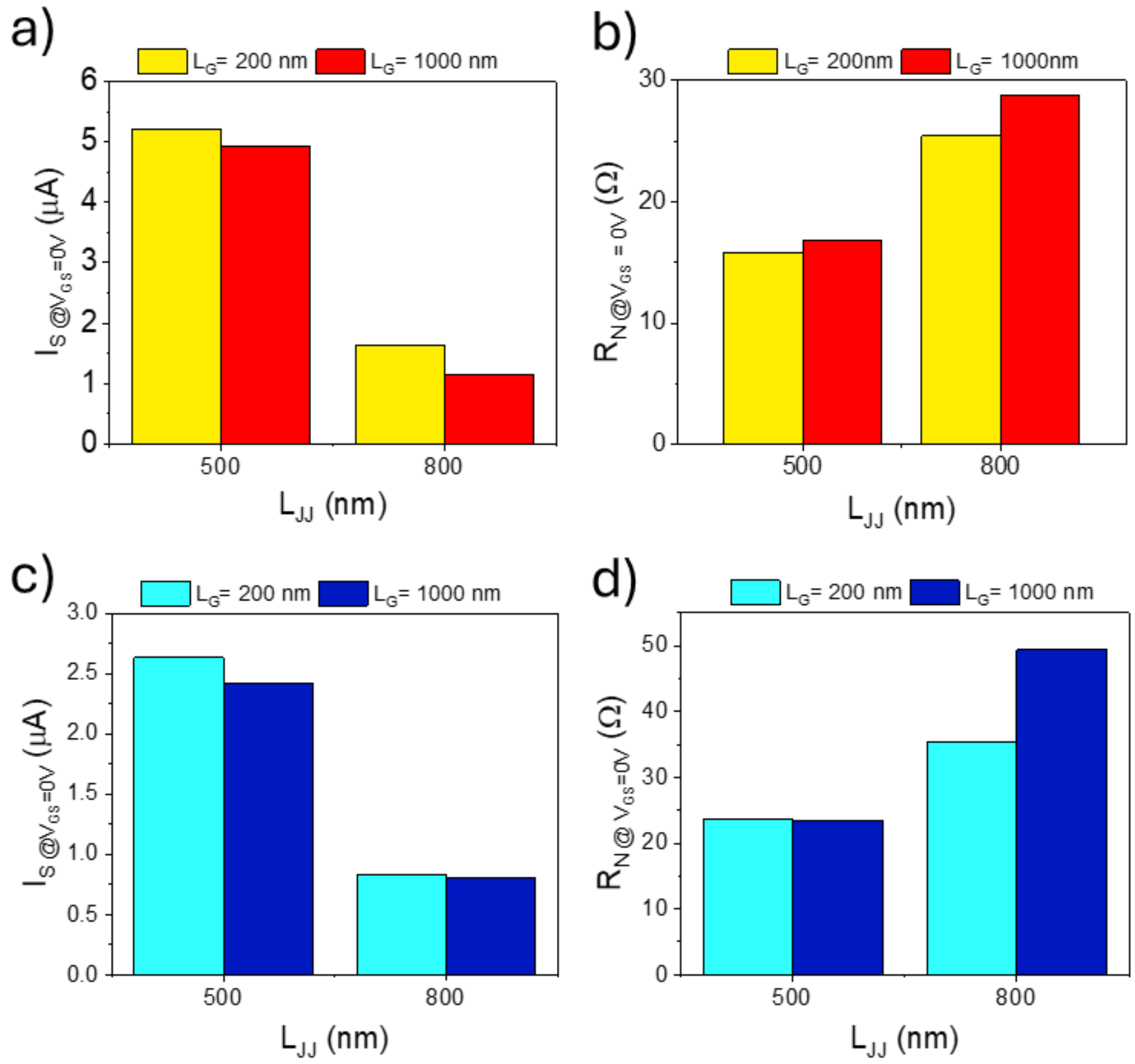

**Figure S3: Zero-gate-voltage electrical properties of InAsOI-based JoFETs.** a,b) Switching current (a) and normal-state resistance (b) of $HfO_2$-JoFETs with different gate lengths ($L_G$) and interelectrode separations ($L_{JJ}$). c,d) Switching current (c) and normal state resistance (d) of $Al_2O_3$-JoFETs with different gate lengths and interelectrode separations.

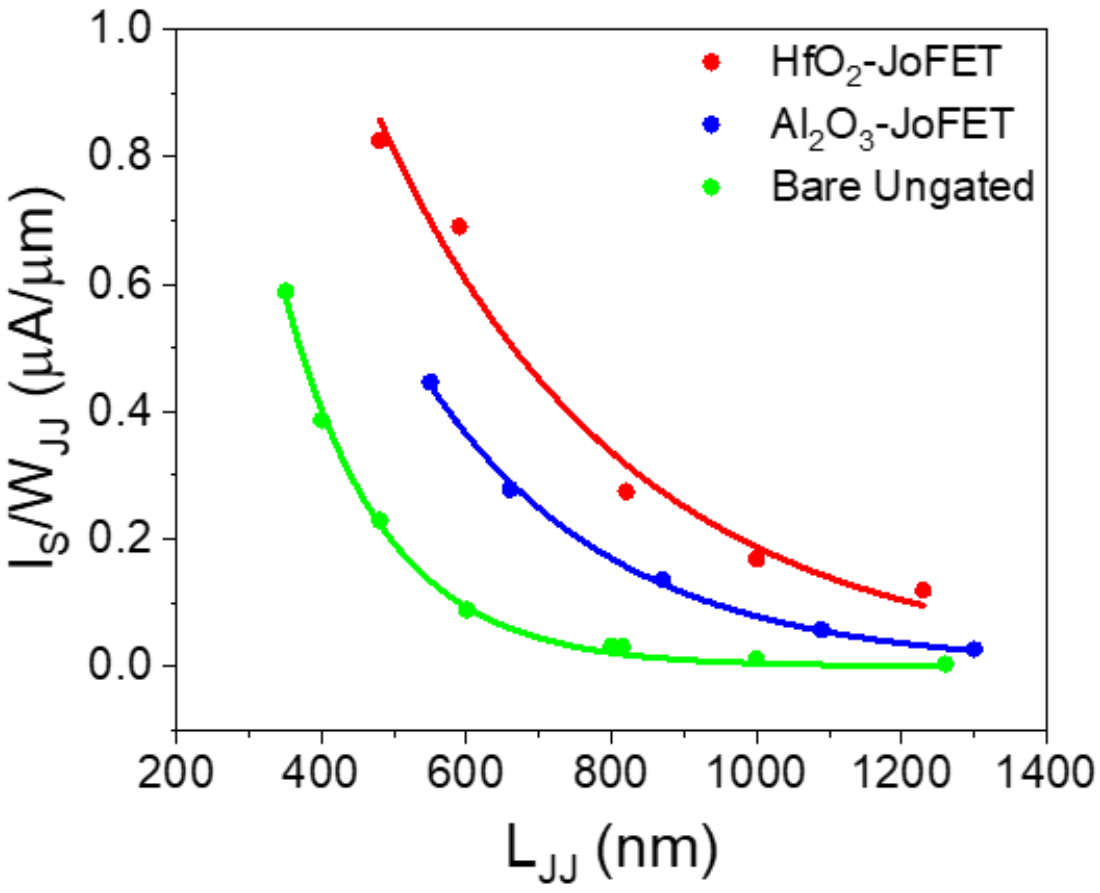

**Figure S4: Zero-gate-voltage switching current density of InAsOI-based JoFETs with different interelectrode separations and gate insulators.** Exponential decay fitting curves were obtained using the fitting function $\frac{I_S}{W_{JJ}} = \frac{I_S}{W_{JJ}}|_{L=0nm} \times e^{-\frac{L}{\tau}}$.

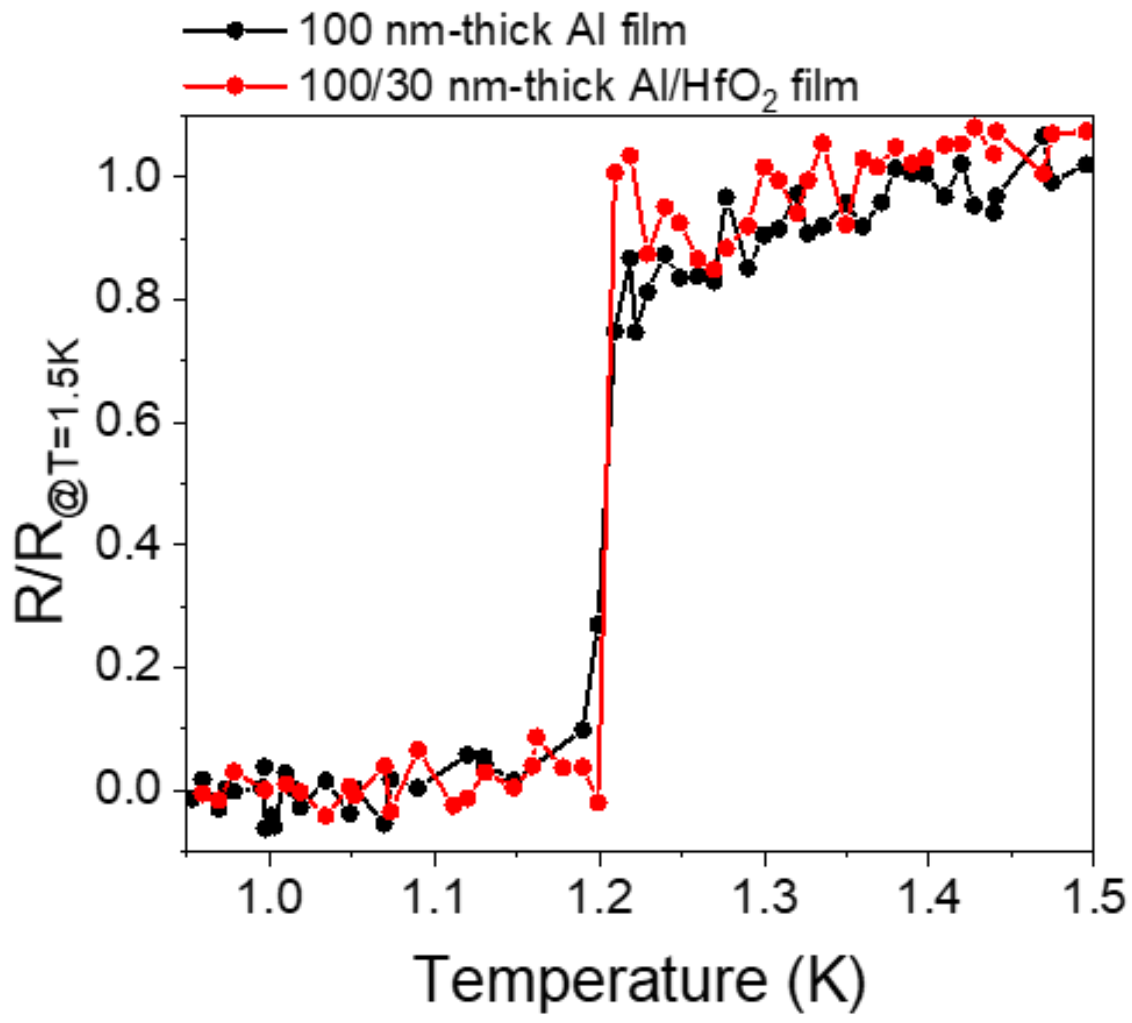

**Figure S5: Temperature-dependent behavior of a 100-nn-thick Al film and a 30/100-nn-thick $HfO_2$/Al film deposited on the InAsOI platform normalized with respect to the value at 1.5 K.** Films were produced in a different experimental run compared to that of the JoFETs.

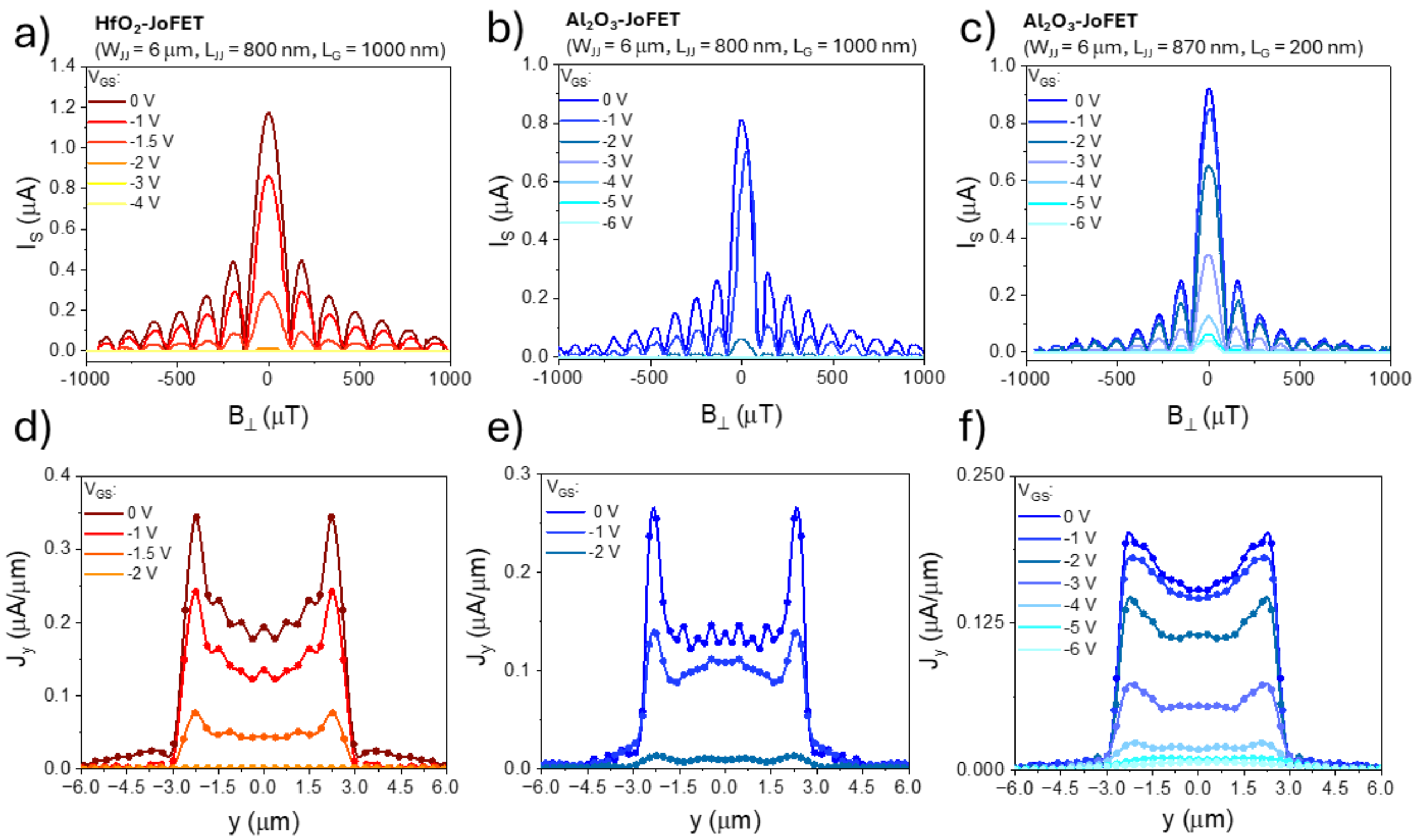

**Figure S6: Out-of-plane magnetic field-dependent behavior of InAsOI-based JoFETs.** a,b,c) Gate-dependent switching current vs. out-of-plane magnetic field characteristic for a JoFET featuring $W_{JJ}$ = 6 μm, $L_{JJ}$ = 800 nm, and $L_G$= 1000 nm, with $HfO_2$ (a) or $Al_2O_3$ (b) as gate insulator and a JoFET featuring $W_{JJ}$ = 6 μm, $L_{JJ}$ = 870 nm, and $L_G$= 200 nm, with $Al_2O_3$ as gate insulator (c). d,e,f) Supercurrent density vs. width position characteristic for a JoFET featuring $W_{JJ}$ = 6 μm, $L_{JJ}$ = 800 nm, and $L_G$= 1000 nm, with $HfO_2$ (d) or $Al_2O_3$ (e) as gate insulator and a JoFET featuring $W_{JJ}$ = 6 μm, $L_{JJ}$ = 870 nm, and $L_G$= 200 nm, with $Al_2O_3$ as gate insulator (f). The measurement temperature is 50 mK.

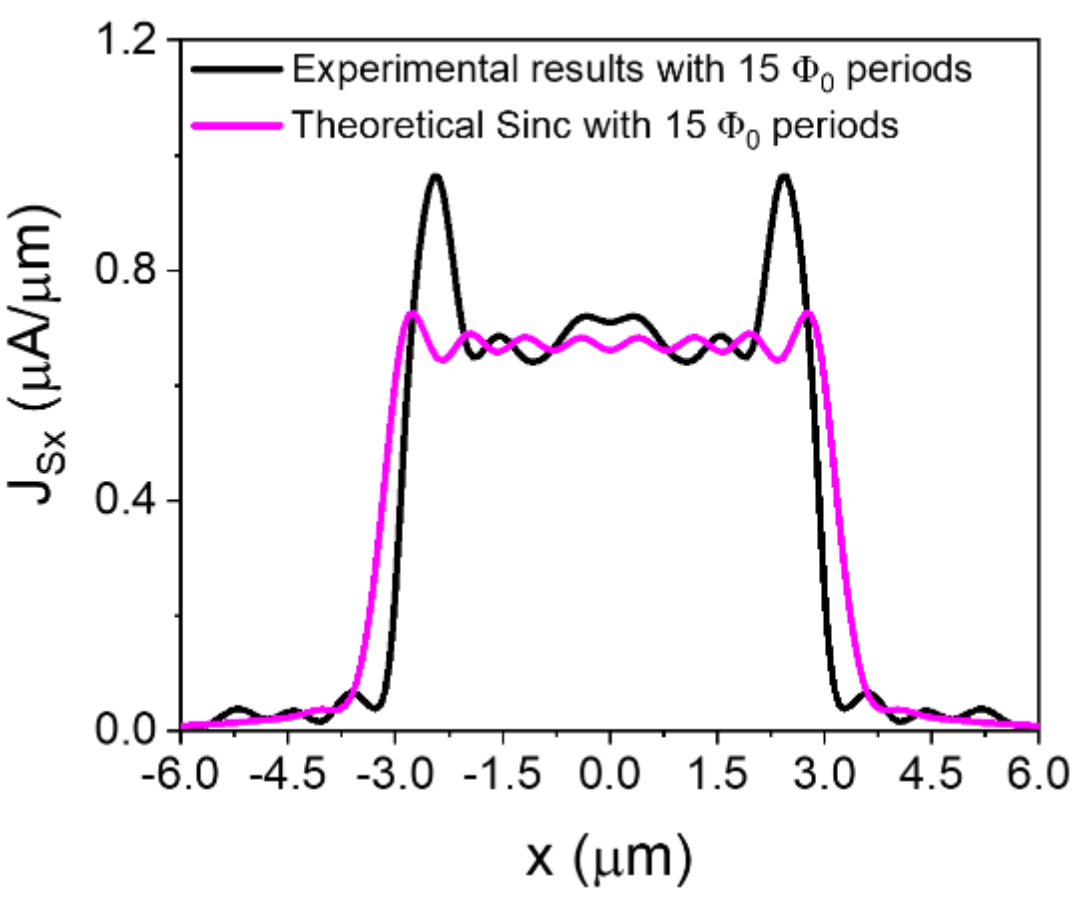

**Figure S7: InAsOI-based JoFET supercurrent density distribution vs. theoretical Sinc-function.** Comparison of the computed supercurrent density distribution for a $HfO_2$-JoFET with $W_{JJ}$ = 6 μm, $L_{JJ}$ = 500 nm, and $L_G$= 1000 nm (black curve) and for a theoretical Sinc-function employing $I_S|_{B_\perp=0T}$ and $W_{JJ}$ equal to that of the $HfO_2$-JoFET (magenta curve). Both curves are obtained from a Fraunhofer-like pattern having 15 $\Phi_0$ periods.

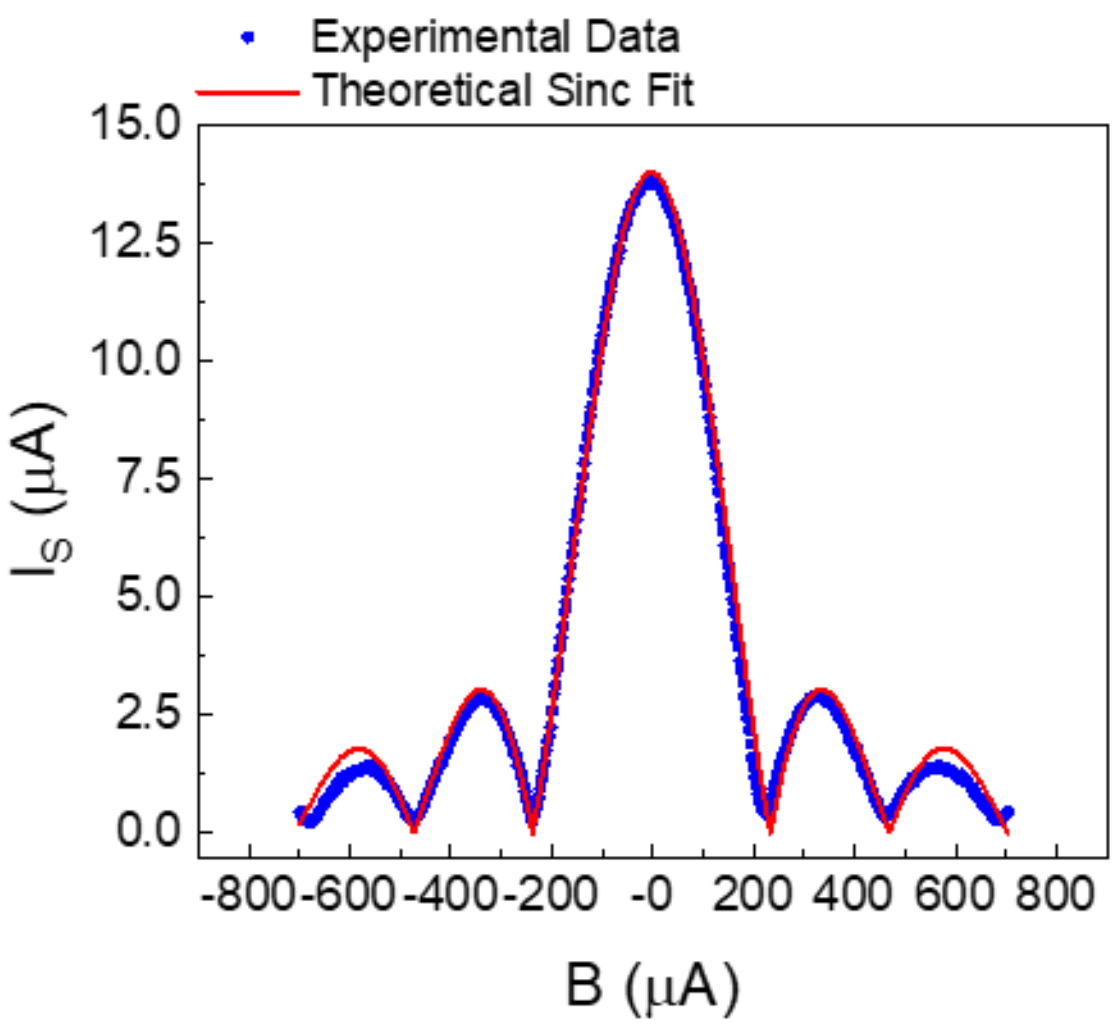

**Figure S8: Out-of-plane magnetic field-dependent behavior of InAsOI-based Josephson Junction with $L_{JJ}$ = 550 nm, $W_{JJ}$ = 6 μm, and superconducting leads length of 500 nm reproduced from [1].** The measurement temperature is 50 mK. The plot reports the experimental switching currents as a function of the out-of-plane magnetic field superimposed with a theoretical sinc fit, showing an excellent agreement between experimental and theoretical data.

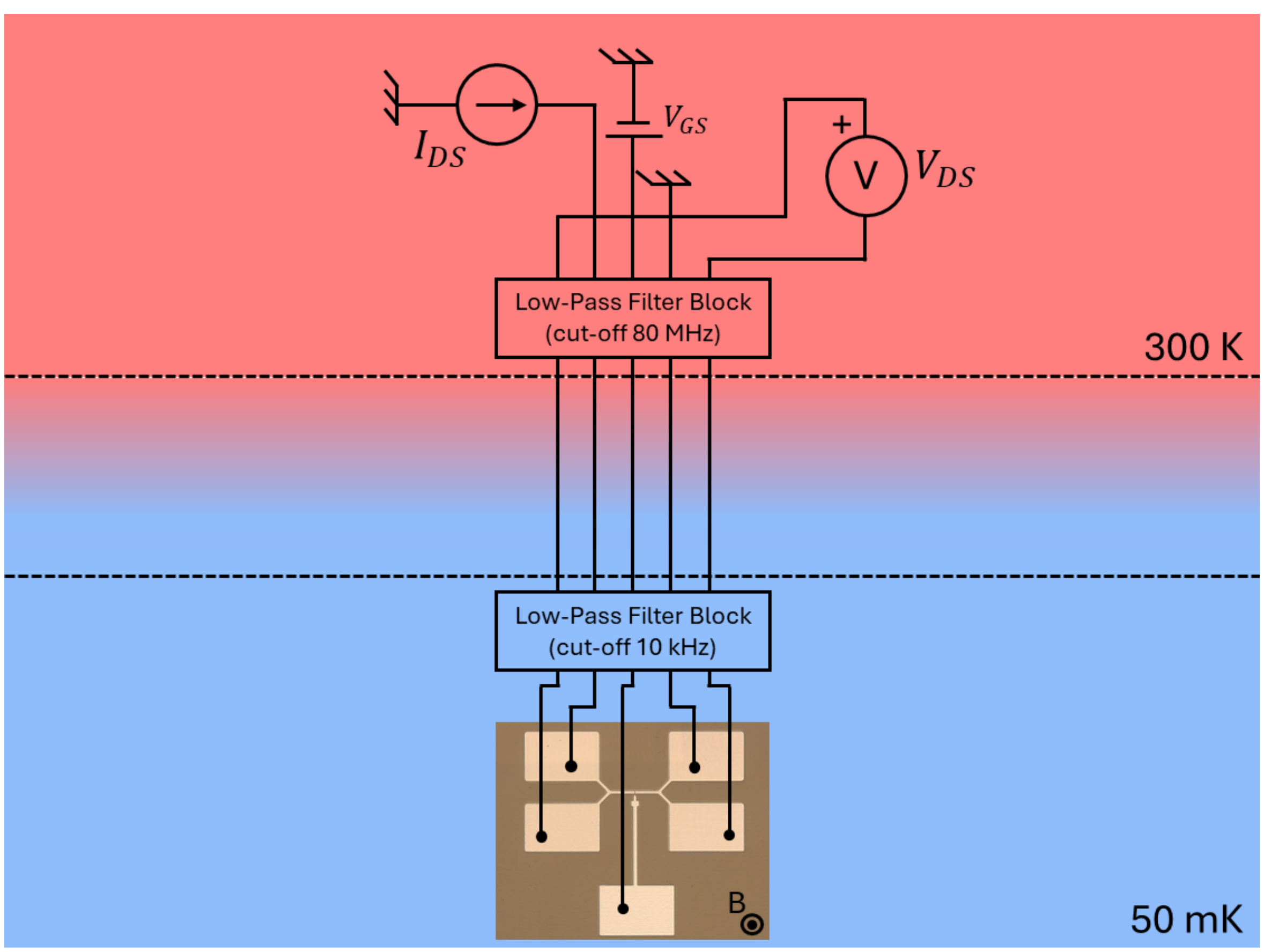

**Figure S9: DC measurement setup used to characterize the JoFETs in a dilution cryostat equipped with a z-axis superconducting magnet.**

## 2. Materials and Manufacturing Methods

### 2.1 Materials and Chemicals

GaAs wafers (2'' diameter, (100) orientation, $\rho=5.9\times10^{7}$ $\Omega\times$cm) were purchased from Wafer Technology LTD. Materials used for the Molecular Beam Epitaxy growth (Gallium 7N5, Aluminum 6N5, Indium 7N5, and Arsenic 7N5) were purchased from Azelis S.A.. Acetone (ACE, ULSI electric grade, MicroChemicals), 2-propanol (IPA, ULSI electric grade, MicroChemicals), S1805 G2 Positive Photoresist (S1805, Microposit, positive photoresist), AR-P 679.04 (AllResist, positive e-beam resist), MF319 Developer (MF319, Microposit), AR 600-56 Developer (AR 600-56, AllResist), AR600-71 (AllResist, remover for photo- and e-beam resist), Aluminum Etchant Type D (Transene), Phosphoric acid ($H_3PO_4$, Sigma Aldrich, semiconductor grade ≥85% in water), Hydrogen peroxide ($H_2O_2$, Carlo Erba Reagents, RSE-For electronic use-Stabilized, 30% in water), Nitrogen ($N_2$, 5.0, Nippon Gases) was provided by the Clean Room Facility of the National Enterprise for nanoScience and nanotechnology (NEST, Pisa, Italy). Diammonium sulfide ($(NH_4)_2S$, Carlo Erba Reagents, 20% in water) was provided by the Chemical Lab Facility of the NEST. Sulfur pieces (S, Alfa Aesar, 99.999% pure) were purchased from Carlo Erba Reagents S.r.l. Aluminum pellets (99.999% pure) were purchased from Kurt J. Lesker Company. Aqueous solutions were prepared using deionized water (DIW, 15.0 M$\Omega\times$cm) filtered by Elix® (Merck Millipore) provided by the Clean Room Facility of the NEST.

### 2.2 InAsOI Heterostructure Growth via Molecular Beam Epitaxy

InAsOI was grown on semi-insulating GaAs (100) substrates using solid-source Molecular Beam Epitaxy (MBE, Compact 21 DZ, Riber). Starting from the GaAs substrate, the sequence of the layer structure includes a 200 nm-thick GaAs layer, a 200 nm-thick GaAs/AlGaAs superlattice, a 200 nm-thick GaAs layer, a 1.250 µm-thick step-graded $In_XAl_{1-X}As$ metamorphic buffer layer (with X increasing from 0.15 to 0.81), a 400 nm-thick $In_{0.84}Al_{0.16}As$ overshoot layer, and a 100 nm-thick InAs layer.

The GaAs layer and the GaAs/AlGaAs superlattice below the $In_XAl_{1-X}As$ buffer layer are grown to planarize the starting GaAs surface and to reduce surface roughness caused by the oxide desorption process. The metamorphic buffer layer consists of two regions with different misfit gradients df/dt. The first $In_XAl_{1-X}As$ region is composed of twelve 50 nm-thick layers with X ramping from 0.15 to 0.58, corresponding to a misfit gradient of 5.1% $\mu m^{-1}$. The second $In_XAl_{1-X}As$ region is composed of twelve 50 nm-thick layers with X ramping from 0.58 to 0.81, corresponding to a misfit gradient around 3.1% $\mu m^{-1}$. The Al flux was kept constant during the buffer layer growth, while the In cell temperature was increased at each step without growth interruptions. Therefore, the In concentration of the buffer is continuously changing with two different slopes. The $In_XAl_{1-X}As$ buffer layer and InAs epilayer were grown at optimized substrate temperatures of 320 °C ± 5 °C and 480 ± 5 °C, respectively. The As flux was adjusted during the growth to keep a constant group V/III beam flux ratio of 8 throughout the growth. A growth rate of 0.96 μm/h was used to grow the thin InAs layer [2,3].

### 2.3 InAsOI-Based Josephson Field Effect Transistor Fabrication

In the following, all wetting steps were performed in cleaned glass Beckers using stainless steel tweezers provided with carbon tips. Teflon-coated tweezers were used for all the steps requiring acid or base solutions. The Julabo TW2 was used to head the solution to a specific temperature.

*Superconductor Deposition on InAsOI*

InAsOI substrates were cut into square samples (7×7 mm×mm) and sonicated in ACE and IPA for 5 min to remove GaAs dusts. The air-exposed InAs surface was etched from native InAs oxide ($InAsO_X$) and passivated with S-termination by dipping the InAsOI samples in a $(NH_4)_2S_X$ solution (290 mM $(NH_4)_2S$ and 300 mM S in DIW) at 45°C for 90 s. The S-terminated InAsOI samples were then rinsed twice in DIW for 30 s and immediately loaded (~ 90 s exposure time in the air) into the load-lock vacuum chamber of an e-beam evaporator (acceleration voltage 7 kV). Samples were transferred into the deposition chamber, where a 100-nm-thick Al layer was deposited at a rate of 2 A/s at a residual chamber pressure of 1E-8 ÷ 5E-9 Torr.

*InAsOI MESA Fabrication*

After Al deposition, a layer of S1805 positive photoresist was spin-coated at 5000 RPM for 60 s (spin coating acceleration of 5000 RPM/s) and soft-baked at 115 °C for 60 s. The resist was then exposed via direct writing UV lithography (UVL, DMO, ML3 laser writer, λ=385 nm) with a dose of 60 $mJcm^{-2}$, resolution of 0.6 μm, high exposure quality, and laser-assisted real-time focus correction to define the MESA geometry. Unless otherwise stated, all the rinsing steps were performed at room temperature (RT, 21 °C). The UV-exposed samples were developed in MF319 for 45 s with soft agitation to remove exposed photoresist, then rinsed in DIW for 30 s to stop the development and dried with $N_2$. The exposed Al layer was removed by dipping the sample in Al Etchant Type D at 40 °C for 65 seconds with soft agitation, then rinsed in DIW for 30 seconds to stop the etching and dried with $N_2$. The exposed InAs epilayer was etched by dipping the samples in a $H_3PO_4$:$H_2O_2$ solution (348 mM $H_3PO_4$, 305 mM $H_2O_2$ in DIW) for 60 s with soft agitation, then rinsed in DIW for 30 s to stop the etching and dried with $N_2$. Eventually, the photoresist was removed by rinsing the InAsOI samples in ACE at 60 °C for 5 minutes and IPA for 60 s, then dried with $N_2$. At the end of this step, the width (W) of the JoFET was set to 6 μm.

*Markers Deposition for Aligned Steps*

After MESA fabrication, a layer of AR-P 679.04 positive e-beam resist was spin-coated at 4000 RPM for 60 s (spin coating acceleration of 10000 RPM/s) and soft-baked at 160 °C for 60 s. The resist was then exposed via UVL-marker-aligned e-beam lithography (EBL, ZEISS, Ultra Plus) with a dose of 350 $\mu Ccm^{-2}$, voltage acceleration of 30 kV, aperture of 7.5 or 120 μm, line step size of 1 nm or 200 nm, to define EBL-markers for the next alignment steps. The electron-exposed samples were developed in AR 600-56 for 90 s with soft agitation to remove the exposed e-beam resist, then rinsed in IPA for 30 s to stop the development and dried with $N_2$. The samples were loaded into a thermal evaporator (Sistec prototype) where a 10/50-nm-thick Ti/Au bilayer was deposited at a rate of 1 A/s at a residual chamber pressure of 2E-6 mbar. The deposited film was lifted-off in ACE at 70 °C for 5 min with strong agitation, then rinsed in IPA for 60 s, and dried with $N_2$.

*Aligned Josephson Junction Fabrication*

After EBL marker definition, a layer of AR-P 679.04 positive e-beam resist was spin-coated at 4000 RPM for 60 s (spin coating acceleration of 10000 RPM/s) and soft-baked at 160 °C for 60 s. The resist was then exposed via marker-aligned EBL (ZEISS, Ultra Plus) with a dose of 350 $\mu Ccm^{-2}$, voltage acceleration of 30 kV, aperture of 7.5 μm, and line step size of 1 nm to define the Josephson junction length ($L_{JJ}$). The electron-exposed samples were developed in AR 600-56 for 90 s with soft agitation to remove the exposed e-beam resist, then rinsed in IPA for 30 s to stop the development and dried with $N_2$. Subsequently, the exposed Al layer was removed by dipping the sample in Al Etchant Type D at 40 °C for 65 seconds with soft agitation, then rinsed in DIW for 30 seconds to stop the etching and dried with $N_2$. Eventually, the e-beam resist was removed by rinsing the InAsOI samples in ACE at 70 °C for 5 min, IPA for 60 s, and dried with $N_2$. At the end of this step, we achieved $L_{JJ}$ ranging from 500 to 1250 nm.

*Gate Insulator Deposition*

Samples were loaded into the vacuum chamber of an Atomic Layer Deposition system (ALD, Oxford Instruments, OpAL) where the high-*k* dielectric was uniformly grown at 130 °C. We deposited single layers of $HfO_2$ or $Al_2O_3$. For $HfO_2$, Tetrakis(ethylmethylamino)hafnium (TEMAH) and $H_2O$ were used as oxide precursors, while Trimethylaluminum (TMAl) and $H_2O$ were involved in the $Al_2O_3$ growth. In both the cases, Ar was used as carrier gas. Ar bubbling was also involved to increase the volatility of TEMAH. After reaching a base pressure of ~3-4 mTorr, the chamber pressure was increased to ~360 mTorr injecting Ar. For both the insulators, the deposition process follows 4 steps: (i) main precursor dose, (ii) main precursor purge, (iii) $H_2O$ dose, and (iv) $H_2O$ purge.

For the $HfO_2$ deposition:

(i) TEMAH dose: TEMAH valve on; Ar bubbler: 250 sccm; Ar purge: 10 sccm; step duration: 0.9 s.

(ii) TEMAH: purge: TEMAH valve off; Ar purge: 250 sccm; step duration: 110 s.

(iii) $H_2O$ dose: $H_2O$ valve on; Ar purge: 10 sccm; step duration: 0.03 s.

(iv) $H_2O$ purge: $H_2O$ valve off; Ar purge: 250 sccm; step duration: 90 s.

For the $Al_2O_3$ deposition:

(i) TMAl dose: TMAl valve on; Ar purge: 50 sccm; step duration: 0.03 s.

(ii) TMAl purge: TMAl valve off; Ar purge: 250 sccm; step duration: 110 s.

(iii) $H_2O$ dose: $H_2O$ valve on; Ar purge: 5 sccm; step duration: 0.025 s.

(iv) $H_2O$ purge: $H_2O$ valve off; Ar purge: 250 sccm; step duration: 90 s.

We performed 250 ALD cycles to achieve a total insulator thickness of ~31 nm [4]. Eventually, the chamber was pumped to reach a base pressure of ~3-4 mTorr and then vented using $N_2$. The entire process takes ~16 h.

*Aligned Metal Gate Deposition*

After gate insulator deposition, a layer of AR-P 679.04 positive e-beam resist was spin-coated at 4000 RPM for 60 s (spin coating acceleration of 10000 RPM/s) and soft-baked at 160 °C for 60 s. The resist was then exposed via marker-aligned EBL (ZEISS, Ultra Plus) with a dose of 350 $\mu Ccm^{-2}$, voltage acceleration of 30 kV, aperture of 7.5 or 120 μm, line step size of 1 or 200 nm, to define the JoFET gate length ($L_G$). The electron-exposed samples were developed in AR 600-56 for 90 s with soft agitation to remove the exposed e-beam resist, then rinsed in IPA for 30 s to stop the development and dried with $N_2$. Samples were mounted into an e-beam evaporator (acceleration voltage 7 kV), where a 6/60-nm-thick Ti/Al bilayer was deposited at a rate of 0.5/2 A/s with a tilt angle of 53° at a residual chamber pressure of 1E-8 ÷ 1E-9 Torr. The deposited film was lifted-off in AR600-71 at 80 °C for 5 min with strong agitation, then rinsed in IPA for 30 s, and dried with $N_2$.

### 2.4 Al and $HfO_2$/Al Films Deposition

InAsOI substrates were cut into square samples (7×7 mm×mm) and sonicated in ACE and IPA for 5 min to remove GaAs dusts. The as-cleaned InAsOI samples were loaded into the load-lock vacuum chamber of an e-beam evaporator (acceleration voltage 7 kV). Samples were transferred into the deposition chamber, where a 100-nm-thick Al layer was deposited at a rate of 2 A/s at a residual chamber pressure of 1E-8 ÷ 5E-9 Torr.

Samples were then loaded into the vacuum chamber of an ALD (Oxford Instruments, OpAL) where $HfO_2$ was uniformly grown at 130 °C using Tetrakis(ethylmethylamino)hafnium (TEMAH) and $H_2O$ used as oxide precursors. Ar was used as carrier gas and Ar bubbling was also involved to increase the volatility of TEMAH. After reaching a base pressure of ~3-4 mTorr, the chamber pressure was increased to ~360 mTorr injecting Ar. The deposition process follows 4 steps:

(i) TEMAH dose: TEMAH valve on; Ar bubbler: 250 sccm; Ar purge: 10 sccm; step duration: 0.9 s.

(ii) TEMAH: purge: TEMAH valve off; Ar purge: 250 sccm; step duration: 110 s.

(iii) $H_2O$ dose: $H_2O$ valve on; Ar purge: 10 sccm; step duration: 0.03 s.

(iv) $H_2O$ purge: $H_2O$ valve off; Ar purge: 250 sccm; step duration: 90 s.

We performed 250 ALD cycles to achieve a total insulator thickness of ~31 nm [4]. Eventually, the chamber was pumped to reach a base pressure of ~3-4 mTorr and then vented using $N_2$. The entire process takes ~16 h.

### 2.5 Sample Bonding via Wire Wedge Bonding

All the fabricated samples were provided with bonding pads ranging from 150 × 150 to 200 × 200 μm×μm and then used to connect the device with the chip carrier. Samples were glued using a small drop of AR-P 679.04, then left dry at RT for 1 hour on a 24-pin dual-in-line (DIL) chip carrier. Samples were bonded via wire wedge bonding (MP iBond5000 Wedge) using an Al/Si wire (1%, 25 μm wire diameter), leaving the user-bonder and the DIL chip carrier electrically connected to the ground.

## 3. Characterization Methods

### 3.1 Morphological Characterization

***via Scanning Electron Microscopy***

Top view morphological characterization of JJs was carried out via scanning electron microscopy (SEM, ZEISS Merlin) with 5 kV acceleration voltage, 178 pA filament current, back scattered electron relevator, at different magnifications (2.5k and 50k).

***via Optical Microscopy***

Optical microscopy (Leica, DM8000 M, provided with LEICA MC190 HD camera) was used to verify all the steps without photoresist. An optical microscope (Nikon, Eclipse ME600, provided with Nikon TV Lens C-0.6× and a UV filter) was used to evaluate all the steps involving the photoresist.

Specifically, all the measured JoFETs were morphologically characterized before the electrical measurement to ensure the absence of imperfections related to the manufacturing process, namely, (i) microscopic defects induced by the MBE growth, (ii) incomplete interelectrode separation after the Al wet etching process, (iii) irregular mesa edges related to the InAs wet etching process, (iv) non-conformal insulator coating after the ALD process, and (v) Ti/Al flakes residual related to the gate electrode lift-off process.

### 3.2 DC Electrical Characterization of InAsOI-Based Josephson Field Effect Transistors

Electrical characterization of JoFETs was carried out by measuring 4-wires V-I curves at 50 mK. Out-of-plane magnetic field ($B_{\perp}$) was used to maximize the switching current at 0 gate voltage, then maintained for all the characterization. The sample was mounted in contact with the mixing chamber (MC) plate of the Leiden CF-CS81-1400 cryostat. Electrical configuration of the measurement setup is shown in Figure S9. Source-drain current ($I_{DS}$) was injected applying an increasing DC voltage (Voltage Source, YOKOGAWA GS200) over an input series resistor (R= 1 MΩ) at least 100 times larger than the total resistance of the remaining measurement setup. The voltage drop across the probe contacts ($V_{DS}$) was amplified (Voltage Amplifier, DL Instruments 1201, Gain = 10k, High pass filter = DC, Low pass filter = 100 Hz) and read (Multimeter, Agilent, 34410A, NPLC = 2). The gate-source voltage ($V_{GS}$) was changed between 0 and -6 V with a step size of 0.5 V (SMU, Keithley, 2400).

The switching current ($I_S$) was estimated as the last applied current in the V-I curve before reading a voltage drop different from the noise floor, while the normal state resistance ($R_N$) was evaluated as the angular coefficient of the V-I linear best fitting curve for $I>I_S$.

Gate current leakage was evaluated measuring the gate I-V curve (with the source terminal grounded and the other terminals left open) upon application of an increasing voltage (SMU, Keithley, 2400, absolute voltage step = 100 mV, step delay = 1 s) and collecting the flowing current (TIA, FEMTO, DDPCA-300, gain = $10^{10}$ V/A, rise time = fast; Multimeter, Agilent, 34410A, NPLC = 2).

V-I curves were measured to collect the Fraunhofer interference patterns, changing $B_\perp$ from -1 to 1 mT with a step size of 10 μT, from which $I_S$ was estimated.

The resistance vs. temperature (R vs. T) behavior of the JoFETs was acquired by measuring the differential resistance of the JoFETs during the cryostat cool down starting from 1.5 K. A digital lock-in amplifier (NF Corporation, LI 5640, oscillator voltage RMS = 0.05 V, oscillator voltage frequency = 17 Hz, integration time = 100 ms, sensitivity = 100 mV) was used to provide an AC source-drain current and measure the differential resistance of the JoFET. The lock-in amplifier oscillator voltage was applied across a series resistor (R=100 kΩ) at least 100 times larger than the total resistance of the remaining measurement setup to apply the AC source-drain current. The AC voltage drop across the probe contacts was amplified (DL Instruments 1201, Gain = 10k, High pass filter = DC, Low pass filter = 100 Hz or 30 Hz) and read by the lock-in amplifier to extrapolate the JJ differential resistance. Then, the data output voltage of the lock-in amplifier has been read (34401 Multimeter, NPLC = 0.1) and recorded.

The resistance vs. temperature (R vs. T) behavior of the Al and $HfO_2$/Al films was acquired by measuring the 4-wire voltage drop across the films during the cryostat cool down starting from 1.5 K. A DC current of 100 μA was injected applying a DC voltage of 10 V (Voltage Source, YOKOGAWA GS200) over an input series resistor (R= 100 kΩ) at least 100 times larger than the total resistance of the remaining measurement setup. The DC voltage drop across the probe contacts was amplified (DL Instruments 1201, Gain = 10k, High pass filter = DC, Low pass filter = 100 Hz or 30 Hz), read (34401 Multimeter, NPLC = 0.1), and recorded.

### 3.3 Supercurrent Density Estimation from Fraunhofer Interference Patterns

The switching current obtained from the Fraunhofer interference pattern ($I_S^{MAX}$) is a function of the effective out-of-plane magnetic field $B_{eff} = B \times \alpha$, where $\alpha > 1$ is the focusing factor. $I_S^{MAX}(B_{eff})$ can be represented as a function of $\beta = f(B_{eff}) = \frac{2\pi}{\Phi_0} L_{eff} \times B_{eff}$, where $L_{eff}$ is the JJ effective length ($L_{eff} > L_{JJ}$, $L_{JJ}$ is the interelectrode separation [1]) and $\Phi_0$ is the flux quantum. Then, $I_S^{MAX}(\beta)$ is the absolute value of the Fourier Transform of the supercurrent density $\left(J_S(x)\right)$, namely $I_S^{MAX}(\beta) = |g(\beta)| = \left|F\left(J_S(x)\right)\right| = \left|\int_{-\infty}^{+\infty} J_S(x) e^{-i\beta x} dx\right|$, from which $J_S(x)$ can be obtained [5,6].

Since the measured $I_S^{MAX}$ has some non-zero minima, in the general case, we consider $g(\beta) = g_E(\beta) + i g_O(\beta)$, where $g_E(\beta)$ and $g_O(\beta)$ are the even and the odd parts, respectively. $g_E(\beta)$ is retrieved by multiplying $I_S^{MAX}(\beta)$ by a flipping function that alternately flips the lobes to obtain a “sinc-like” function. In the $I_S^{MAX}(\beta)$ minima, we suppose $g_O(\beta)$ dominates; $g_O(\beta)$ is retrieved by interpolating between the $I_S^{MAX}(\beta)$ minima and flipping sign between lobes [5,7]. We neglect $g_O(\beta)$ since it simply introduces a very small asymmetry in the current density distributions, which is not detrimental for the purposes of our work.

In the end, the Inverse Fast Fourier Transform (iFFT) of $g(\beta)$ is performed to get $J_S(x)$. The $x$-range ($\Delta_x$) is calculated as $\Delta_x = \frac{2\pi}{\delta_\beta}$, where $\delta_\beta$ is the $\beta$-spacing. No window function was used in the iFFT.

Since the contribution of the focusing factor $\alpha$ of the magnetic field and the elongation of the JJ due to diffusive phenomena cannot be factorized in our approach, we defined a number that takes into account both these effects $\gamma = L_{eff} \times B_{eff} / (L_{JJ} \times B)$. To determine $\gamma$, we fixed the JJ width ($W$) measured via SEM and we changed $\beta$ to make the supercurrent density distribution x-axis match $W$. Table S1 reports $\gamma$ values for the measured JoFETs.

**Table S1: Values of $\gamma$ for the measured InAsOI-based JoFETs.**

| JoFET Characteristics | | | | $\gamma$ |
|---|---|---|---|---|
| $L_{JJ}$ [nm] | $W_{JJ}$ [µm] | $L_G$ [nm] | Gate Insulator | |
| 480 | 6.4 | 1000 | $HfO_2$ | 5.5 |
| 800 | 5.5 | 1000 | $HfO_2$ | 3.5 |
| 480 | 5.9 | 1000 | $Al_2O_3$ | 6.0 |
| 800 | 6.1 | 1000 | $Al_2O_3$ | 4.6 |
| 870 | 5.9 | 200 | $Al_2O_3$ | 4.0 |

### 3.4 Supercurrent Density Estimation from Fraunhofer Interference Patterns: Python Script

```
# Author:           Laura Borgongino
# Affiliation:      NEST, Istituto Nanoscienze-CNR and Scuola Normale Superiore,
#                   Pisa, Italy
# Version:          1.0
# Date:             July 2024
# Description:      This code is made for the analysis of the supercurrent
                    distribution from the Fraunhofer pattern

import numpy as np
import pylab as plt
from scipy.signal import find_peaks
from scipy.fftpack import ifft, ifftshift

# Select the file
file_name="Pattern_F.txt"

# Set the length of the Josephson Junction
ljj = 0.48 #um

# Set the parameter gamma (due to focusing and effective length)
gamma = 6

# Read data from txt file
data = np.loadtxt(file_name, unpack=True)
Bz = data[0] # Magnetic field
Is = data[1] # Switching current

# With approx Is(beta)=|C(beta)|. Reconstruct the function w/o modulus:
# 1) Find minima of the pattern
# 2) Define a flipping function
# 3) Multiplying for a flipping function at each range

# Find minima of the pattern
```

```python
mins, _ =find_peaks(Is*-1,prominence=0.001,distance=9)

# Define flipping function to get a sinc-like function
def flip(i,x):
    if (i%2)==0:
        return (-1)*x
    else:
        return x
flip_g=np.ones(len(Is))

# Construct the even part
Is_even=Is[0:mins[0]]
for i in range(0,len(mins)-1):
    Is_=flip(i,Is[mins[i]:mins[i+1]])
    Is_even=np.append(Is_even,Is_)
Is_even=np.append(Is_even,Is[mins[len(mins)-1]:])

# Write in terms of beta
phi_0=2.067833848*1e3 #Wb=uT*um^2
beta=gamma*Bz*2*np.pi*ljj/phi_0 # 1/um=uT*um/(uT*um^2)

# Obtain the current distribution via ifft. You need to multiply for the beta-
spacing and the number of points
Jx=len(beta)*np.abs((beta[0]-beta[1])*ifft(Is_even))/(2*np.pi)

# Shift the zero frequency component to the centre of the spectrum
Jx=ifftshift(Jx)

# Obtain the position
Xmax=2*np.pi/np.abs(beta[0]-beta[1])
x=np.linspace(-Xmax/2,Xmax/2,num=len(Jx))

# Plot the current distribution as a function of the position
plt.xlabel("Position (um)")
plt.ylabel("Current density distribution (uA/um)")
plt.plot(x, Jx)
plt.show()
```

### 3.5 InAs Coherence Length and Electron Mean Free Path Estimation

The InAs coherence length ($\xi_0$) is calculated using the following equation:

$$\xi_0 = \sqrt{\frac{\hbar D}{\Delta}} = 651\ nm$$

where D is the diffusion constant of the InAs epilayer, $\Delta = 1.76\ k_B T_c$ is the aluminum superconducting gap, ħ and $k_B$ are the reduced Planck and Boltzmann constant, respectively, and $T_c$ is the aluminum critical temperature [8,9].

D is calculated from the relationship:

$$D = \frac{1}{2} v_F l_e = 0.1\ m^2 s^{-1}$$

where $l_e$ is the mean free path of the InAs layer, and $v_F$ is its Fermi velocity.

$v_F$ is obtained from the formula:

$$v_F = \sqrt{\frac{2E_F}{m^*}} = 1.5 \times 10^6 ms^{-1}$$

where $m^* = 0.023 m_e$ is the effective mass of InAs, with $m_e$ the electron mass, and $E_F$ is the Fermi energy.

Fermi energy depends on the sheet electron density ($n_{2D}$) via the relationship:

$$E_F = \frac{\pi \hbar^2}{m^*} n_{2D} = 2.3 \times 10^{-20} J$$

where we assume that the donors are consistently fully ionized in the conduction band and that the Fermi level is positioned within it.

The InAs electron mean free path was estimated from the following relationship:

$$l_e = \frac{\mu_n m^* v_F}{e} = 131\ nm$$

where $\mu_n$ is the InAs epilayer electron mobility.

## References

(1) Battisti, S.; Simoni, G. De; Braggio, A.; Paghi, A.; Sorba, L.; Giazotto, F. Extremely Weak Sub-Kelvin Electron–Phonon Coupling in InAs on Insulator. *Appl. Phys. Lett.* **2024**, *125* (20). https://doi.org/10.1063/5.0225361.

(2) Arif, O.; Canal, L.; Ferrari, E.; Ferrari, C.; Lazzarini, L.; Nasi, L.; Paghi, A.; Heun, S.; Sorba, L. Influence of an Overshoot Layer on the Morphological, Structural, Strain, and Transport Properties of InAs Quantum Wells. *Nanomaterials* **2024**, *14* (7), 592. https://doi.org/10.3390/nano14070592.

(3) Paghi, A.; Trupiano, G.; De Simoni, G.; Arif, O.; Sorba, L.; Giazotto, F. InAs on Insulator: A New Platform for Cryogenic Hybrid Superconducting Electronics. *Adv. Funct. Mater.* **2025**, *35* (7). https://doi.org/10.1002/adfm.202416957.

(4) Paghi, A.; Battisti, S.; Tortorella, S.; De Simoni, G.; Giazotto, F. Cryogenic Behavior of High-Permittivity Gate Dielectrics: The Impact of the Atomic Layer Deposition Temperature and the Lithography Pattering Method. *Prepr. http//arxiv.org/abs/2407.04501* **2024**.

(5) Hart, S.; Ren, H.; Wagner, T.; Leubner, P.; Mühlbauer, M.; Brüne, C.; Buhmann, H.; Molenkamp, L. W.; Yacoby, A. Induced Superconductivity in the Quantum Spin Hall Edge. *Nat. Phys.* **2014**, *10* (9), 638–643. https://doi.org/10.1038/nphys3036.

(6) Elfeky, B. H.; Lotfizadeh, N.; Schiela, W. F.; Strickland, W. M.; Dartiailh, M.; Sardashti, K.; Hatefipour, M.; Yu, P.; Pankratova, N.; Lee, H.; Manucharyan, V. E.; Shabani, J. Local Control of Supercurrent Density in Epitaxial Planar Josephson Junctions. *Nano Lett.* **2021**, *21* (19), 8274–8280. https://doi.org/10.1021/acs.nanolett.1c02771.

(7) Dynes, R. C.; Fulton, T. A. Supercurrent Density Distribution in Josephson Junctions. *Phys. Rev. B* **1971**, *3* (9), 3015–3023. https://doi.org/10.1103/PhysRevB.3.3015.

(8) Cuevas, J. C.; Bergeret, F. S. Magnetic Interference Patterns and Vortices in Diffusive SNS Junctions. *Phys. Rev. Lett.* **2007**, *99* (21), 217002. https://doi.org/10.1103/PhysRevLett.99.217002.

(9) Bergeret, F. S.; Cuevas, J. C. The Vortex State and Josephson Critical Current of a Diffusive SNS Junction. *J. Low Temp. Phys.* **2008**, *153* (5–6), 304–324. https://doi.org/10.1007/s10909-008-9826-2.